\documentclass[aps,nofootinbib,amsmath,prc,showpacs,groupedaddress,12pt,superscriptaddress,notitlepage]{revtex4-1}
\pdfoutput=1

\usepackage[paperwidth=210mm,paperheight=297mm,centering,hmargin=2.3cm,tmargin=2.8cm,bmargin=3.4cm]{geometry}
\usepackage{amsfonts}
\usepackage{amsmath}
\usepackage{epsfig}
\usepackage{latexsym}
\usepackage{paralist}
\usepackage{fancyhdr}
\usepackage{graphicx}

\numberwithin{equation}{section}

\usepackage[vcentermath]{youngtab}
\usepackage{young}
\usepackage{ytableau}
\usepackage{etex}
\usepackage{braket}
\usepackage{float}
\usepackage{slashed}
\usepackage{xcolor}

\usepackage{soul} 

\usepackage{dcolumn} 

\makeatletter
\@addtoreset{equation}{section}

\makeatother

\def\bea{\begin{eqnarray}} 
\def\eea{\end{eqnarray}}
\def\be{\begin{equation}} 
\def\ee{\end{equation}} 
\def\ba{\begin{array}}
\def\ea{\end{array}}

\def\be{\begin{equation}}
\def\ee{\end{equation}}
\def\bea{\begin{eqnarray}}
\def\eea{\end{eqnarray}}

\let\oldtitle\title
\renewcommand{\title}[1]{\oldtitle{\color{blue}{#1}}}

\let\oldeqref\eqref
\let\oldcite\cite
\renewcommand{\eqref}[1]{{\color{blue}\oldeqref{#1}}}
\renewcommand{\cite}[1]{{\color{blue}\oldcite{#1}}}

\makeatletter
\let\reftagform@=\tagform@
\def\tagform@#1{\maketag@@@{\ignorespaces\textcolor{blue}{(\ignorespaces #1 \unskip\@@italiccorr \ignorespaces)\ignorespaces}}}
\makeatother

\makeatletter
\renewcommand{\p@subsection}{}
\renewcommand{\p@subsubsection}{}
\makeatother

\definecolor{realgreen}{rgb}{0.0, 200, 0.0}
\definecolor{amber}{rgb}{1.0, 0.49, 0.0}
\definecolor{cyellow}{rgb}{1.0, 0.5, 0}
\definecolor{deeppink}{rgb}{1.0, 0.08, 0.58}
\definecolor{cgreen}{rgb}{0., 0.67, 0.13}

\usepackage{mathpazo}
\usepackage{amsmath}

\begin{document}
%%%%%%%%%%%%%%%%%%%%%%%%%%%%%%%%%%%%%%%%%%%%%%%%%%%

\title{Functional RG approach to the Potts model}

\author{Riccardo Ben Al\`i Zinati}
%\email{rbenaliz@sissa.it}
\affiliation{SISSA, via Bonomea 265, 34136 Trieste, Italy}
\affiliation{INFN - Sezione di Trieste, I-34136 Trieste, Italy}

\author{Alessandro Codello}
%\email{codello@cp3-origins.net}
%\affiliation{CP$^3$-Origins, 
%Centre for Cosmology and Particle Physics Phenomenology\\
%University of Southern Denmark,
%Campusvej 55, 5230 Odense M, Denmark}
\affiliation{INFN - Sezione di Bologna, via Irnerio 46, 40126 Bologna, Italy}
\affiliation{ICTP South American Institute for Fundamental Research, IFT-UNESP, S$\tilde{a}$o Paulo, SP Brazil  01440-070}

%%%%%%%%%%%%%%%
\begin{abstract}
%%%%%%%%%%%%%%%
\vspace{3mm}
The critical behavior of the $(n+1)$-states Potts model in $d$-dimensions
is studied with functional renormalization group techniques.
We devise a general method to derive  $\beta$-functions
for continuous values of $d$ and $n$ and we write the flow equation for the effective potential (LPA') when instead $n$ is fixed.
We calculate several critical exponents, which are found to be in good agreement with Monte Carlo simulations and $\epsilon$-expansion results available in the literature.
In particular, we focus on {\tt Percolation} $(n\to0)$ and {\tt Spanning Forest}  
$(n\to-1)$ which are the only non-trivial universality classes in $d=4,5$ and where our methods converge faster.
\end{abstract}

\maketitle

\section{Introduction}

The most important problem of statistical field theory
%, and generally of any field theory, 
is the  classification of universality classes in arbitrary dimension and for a general symmetry group. Universality not only works as an unifying principle that organizes in equivalence classes the large number of definable models, 
%in a given dimension and with a given symmetry,
but crucially highlights and explains which properties, both qualitative and quantitative, can be  observed in experiments and more generally in nature. 
While the most simple cases in which the order parameter field $\varphi$ is single component and is invariant under the group $\mathbb{Z}_2$ have beed studied very deeply and describe an infinite family of unitary\footnote{ In the non-unitary case the $\mathbb{Z}_2$-symmetry is replaced by $\mathcal{PT}$-symmetry \cite{zambellizanusso, Codello0, Codello1}.} universality classes, the first of which is the well known {\tt Ising} universality class %and all the relative multi-critical theories
 %{\tt Tricirtical}, {\tt Tetracritical}, ... 
 (see \cite{Codelloscaling1, Codelloscaling2, Codelloscaling3, Codello1} for an analysis in arbitrary dimension); more general {\it discrete} symmetries have received much less attention and the general properties of their theory spaces are still quite unknown in dimensions higher than two, both qualitatively and quantitatively.
Remaining in the realm of finite groups, the two principal generalizations of the $\mathbb{Z}_2$ Ising model are the $\mathbb{Z}_n$ family of abelian clock models and the $S_{n+1}$ family of Potts models. While the first is always characterizable by a two dimensional order parameter field, the implementation of the permutation group symmetry requires an $n$-component field.

In this work we will begin a general study of the {\tt Potts}$_{q}$ ($q\equiv n+1$) universality classes in arbitrary dimensions by extending the functional RG (FRG) approach to field theories with global $S_q$-symmetry.
This family of universality classes includes a great variety of interesting cases, but in particular we will be interested in the cases of {\tt Percolation} ($q=1$) and {\tt Spanning Forest} ($q=0$), the only non-trivial cases, apart {\tt Ising} ($q=2$), in dimension grater than two.
One of the main virtues of the FRG approach is its simple adaptation to any field theory without any restrictions on $d$. In this respect the method is one of the few analytical tools that are general enough to allow a systematic study of universality and our work is the first of a series intended  at the exploration of field theories with discrete symmetries in $d>2$.

%\subsubsection{Generalities}

Since its introduction in 1952 \cite{Potts1952}, the Potts model
has attracted an increasingly amount of attention stimulating both
theoretical and experimental research: despite a simple definition,
the model exhibits a rich critical behaviour reflecting a very different
class of physical situations.
The three state version of the model can describe the transition of
a liquid crystal from its nematic to its isotropic phase \cite{DeGennes1969, DeGennes1971},
the transition of a cubic crystal into a tetragonal phase \cite{Weger1973},
as well as the deconfinment phase transition in QCD at finite temperature
\cite{Svetitsky1982, Yaffe1982, Delfino2, Delfino3, Delfino5}. The two state version is of course
the Ising model. The Potts model with a single state can describe
the critical behaviour of bond percolation \cite{Fortuin1972, Fortuin1969, Harris1975},
while the limit of zero states is related both to the electrical resistor
network and the spanning forest problem \cite{Baxter1973, Sokal2005}.
The Potts model has also been territory of controversy
and debate: according to Landau's phenomenological theory, the presence of the nonzero third order term in the corresponding Lagrangian implies that it undergoes a first order phase transition in any
dimension \cite{Nienhuis1979, Alexander1974}. This opened the long and entangled
problem on the nature of the phase transition in the Potts model.
Baxter \cite{Baxter1973} proved rigorously in 1973 that in two dimensions
it undergoes a second order phase transition for $\ensuremath{q\leq4}$
and a first order one for $\ensuremath{q>4}$ and still is the only
case known exactly. Despite a satisfactory picture for $q_{c}$ in
$d\geq2$ is still missing, numerical simulations performed mainly
in the 70s \cite{Jensen1979, Blote1979, Hermann1979, Fukugita1989, DeAlcantara1991, Gavai1989}
and RG analysis \cite{Golner1973, ZiaWallace1975, Amit1976, Amit1974, Nienhuis1979, Newman1984, DeAlcantara1980, DeAlcantara1981}
suggest for example that in three dimensions $2<\ensuremath{q_{c}}<3$
while in $d=3$ \texttt{Potts$_{3}$} undergoes a first order phase transition.
(The existence of a critical value of $q_{c}$ should correspond to
a collapse of fixed points in the RG formalism \cite{Newman1984}).
It is worth to mention here that quantum fluctuations can change the nature of the phase transition. This is the case for the quantum phase transition of gapless Dirac  fermions coupled to  a $\mathbb{Z}_3$ symmetric order parameter  within a Gross-Neveu-Yukawa model in $2+1$ dimensions, suitable to describe  the  Kekul\'e transition in honeycomb lattice materials. It has been proven recently from an RG and fRG analysis that quantum fluctuations of the massless Dirac fermions at zero temperature can render the putative first-order character of the transition continuous \cite{Li, Scherer1, Scherer2}.
%
%\subsubsection{Open problems}

The main open problems concerning the {\tt Potts}$_{q}$ family of universality classes are the precise quantitative determination of the critical properties of {\tt Spanning Forest} and {\tt Percolation} in $d\geq3$ and the determination of the critical $q_c$ in $d\geq2$ at which the phase transition ceases to be continuous. 
%, or equivalently the determination of $d_{stf}$ at fixed $q$.
%
%\subsubsection{What we do}
%
In this work we will address mainly the first of the above problems by giving estimates for the critical exponents for {\tt Spanning Forest} and {\tt Percolation}  in $d=4,5$ (and preliminary results in $d=3$) while we will postpone to a future work the question related to the critical $q_c$ separating dis-continuous from continuous phase transitions, for which, in any case, we do the preparatory work obtaining the flow equation for the effective potential (LPA') for the three-states case ($n=2$), the flow of which can, in principle, be used to determine the location of a first order phase transition as a function of $d$.
%\footnote{We can better do $d_c$ at fixed $q$!}   

In section \ref{PFT} we review and generalize the construction of the Potts
field theory, i.e. the field theory for an $n$-component scalar with
discrete global symmetry $S_{n+1}$.
Then in section \ref{FRG} we implement the construction of the exact functional
RG equation for the Potts field theory and develop two approximations:
an algorithm to compute the beta functions
of power interactions for arbitrary $n$ and the LPA' for fixed $n$, both in arbitrary dimension.
In section \ref{Applications} we use the beta functions so derived
to the study of the critical properties of the {\tt Potts}$_{n+1}$
universality classes. After a preliminary study which allows
us to make connection with the $\epsilon$-expansion,
we push our approach to obtain accurate estimates
for the critical exponents in $d=4,5$. 
%where usually only numerical methods have been employed to date.
While in these two dimensions we are able to achieve, respectively,  almost full or full
converging estimates for the critical exponents within the truncation employed, in $d=3$ we obtain only preliminary results.

\section{Potts field theory \label{PFT}}

\subsection{Potts Model}

Originally proposed by Potts \cite{Potts1952} as a generalization of the Ising model, the Potts model consists of a statistical model of interacting spins, where at each site of a lattice there is a variable $\sigma_{i}$ that takes $q$ discrete values, $\sigma_{i}=1,2,\dots,q$. In this model two adjacent spins have an interaction energy given by $J\delta(\sigma_{i},\sigma_{j})$, so that it assumes one value when two nearest-neighbor spins are different and another when they are the same, namely
\begin{equation}
\delta(\sigma_{i},\sigma_{j})=\begin{cases}
1 & \text{if \ensuremath{\sigma_{i}=\sigma_{j}}}\\
0 & \text{if \ensuremath{\sigma_{i}\neq\sigma_{j}}}
\end{cases}
\end{equation}
The model is ferromagnetic when $J>0$ and anti-ferromagnetic when
$J<0$ and the Hamiltonian reads
\begin{equation}
\mathcal{H}=-J\sum_{\langle ij\rangle}\delta(\sigma_{i},\sigma_{j})\,.
\end{equation}
This expression is invariant under the group $S_{q}$ of the permutations
of $q$ objects. It is clear that the nature of the values taken by the spins is completely inessential: instead of the $q$ values listed above, one can consider other $q$ distinct numbers or variables of other nature, for example $q$ different colors. 

\begin{figure}
\begin{centering}
\includegraphics[scale=0.5]{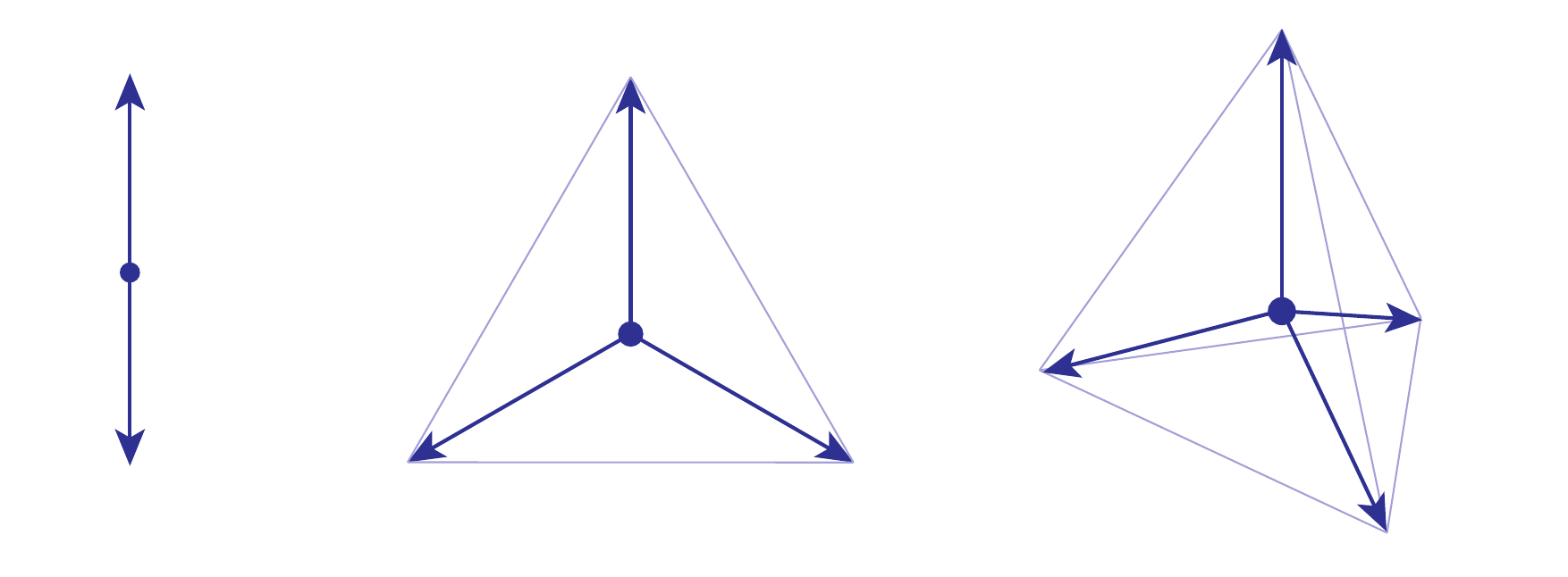}
\par
\end{centering}
\caption{The discrete symmetries characterizing the {\tt Potts}$_{n+1}$ universality classes are those of the $n$-simplex, here shown for $n=1,2,3$. \label{sim}}
\end{figure}

The model can be alternatively formulated to reflect its full symmetry
in a $n=q-1$ dimensional space  \cite{wu1982potts,
ZiaWallace1975}. This is achieved by writing
\begin{equation}
\delta(\alpha,\beta)=\frac{1}{q}\left[1+e^{\alpha}\cdot e^{\beta}\right]\,,
\end{equation}
where $e^{\alpha}$ are the $q$-vectors pointing in the $q$-symmetric
directions of a {\it simplex} in $n=q-1$ dimensions (sometimes referred to as hyper-tetrahedron).
Geometrically  the symmetries of the Potts model are thus those of an $n$-simplex (see Figure~\ref{sim}).

\subsection{Universality classes}

The $q$-states Potts model, in its continuous formulation, describes the universality classes associated to the spontaneous breaking of the permutation symmetry of $q$-colors. Baxter \cite{BaxterB} proved that in two dimensions the transition is continuous for $q\leq 4$. Nevertheless near two dimensions the critical value $q_c(d)$  below which the transition is second order, decreases rapidly as a function of $d$. It is known from a variational RG analysis \cite{Nienhuis1979} that $q_c$ is already lower than three in $d\simeq2.32$ and therefore the  $3$-states Potts model undergoes a first order phase transition in $d=3$. For $q<2$ instead the transition is continuous in all the critical range $2\leq d \leq 6$. 
The obvious question of the meaning of $S_q$ symmetry for a non-integer $q$ arises spontaneously. Long time ago Fortuin and Kasteleyn \cite{Fortuin1972, Fortuin1969} introduced the Random Cluster Model (RCM) as a model for phase transitions and other phenomena in lattice systems, or more generally in systems with a graph structure, where formal use of the symmetry unambiguously leads to final expressions containing $q$ as a parameter which could be varied continuously.

The following is a summary of the present knowledge about the universality classes with $S_q$ symmetry:

\begin{itemize}
\item \texttt{Potts}$_{q}$.  In their seminal paper Fortuin and Kasteleyn  showed that the Potts partition function $Z=\sum_{\left\{\sigma\right\}}\text{e}^{-\mathcal{H}}$ can be written, up to an inessential constant, as
\begin{equation}\label{eq:ZRCM}
Z=\sum_{G \subseteq \mathcal{L}}p^{n_b}(1-p)^{\bar{n}_b}q^{N_c}
\end{equation}
where $G$ is a graph obtained putting $n_b$ bonds on the lattice $\mathcal{L}$, each one with weight $p=1-e^{-J}\in[0,1]$ ($\bar{n}_b$ is the number of absent bonds in $\mathcal{L}$) and $N_c$ is the number of clusters in $G$. The partition function \eqref{eq:ZRCM} defines the RCM. The probability measure for the graph $G$ depends on $q$ through the factor $q^{N_c}$ and it is well defined for any real positive $q$. In
the thermodynamic limit the random cluster model undergoes a phase transition associated to the appearance of a non-zero probability of finding an infinite cluster. By specifying the values of $q$ to $0,1,2,3,\dots$ the model is able to capture at once the phase transition of well known statistical models, the most important of which are listed below.

\item \texttt{\textcolor{black}{Spanning Forest}} \texttt{= Potts$_{0}$}.
%Consider $G=(V,E)$ be finite undirected graph with vertex set $V$
%and edge set $E$.
In combinatorics the multivariate generating polynomial
$Z_{G}(q,p)$ that enumerates the spanning subgraphs (i.e. a subgraph
containing all vertices) of $G$ according to their precise edge content
(with weight $p$ for the edge) and their number of connected
components (with weight $q$ for each component) is called multivariate
Tutte polynomial and is known in statistical physics as the partition
function of the $q$-state Potts model in the form \eqref{eq:ZRCM}. The limit of $q\to0$ with $p/q$ fixed selects the generating polynomial of spanning forests/trees as well as the one of electrical networks introduced long time ago by Kirchhoff \cite{Grimmet, Delfino1}.

\item \texttt{Percolation = Potts$_{1}$}. The limit $q\to 1$ which eliminates the factor $q^{N_c}$ in \eqref{eq:ZRCM}, describes ordinary bond percolation. This formulation, after its introduction, has been revisited and extended to site percolation in 1978 by Wu \cite{Wu1978}. 

\item \texttt{Ising = Potts$_{2}$} thanks to the group isomorphism $\mathbb{Z}_{2}\cong S_{2}$.

\item \texttt{Potts}$_{3}$. 
It is related to the $\mathbb{Z}_3$ model since $S_3\cong\mathbb{Z}_3\times\mathbb{Z}_2$ and in $d=2$ it has the same central charge ($c=\frac{4}{5}$)
as the {\tt Tricritical} universality class.  
The three states version of the model has various connections from nematic to isotropic phase in liquid crystals \cite{DeGennes1969, Alexander1974} to the deconfinment phase transition of mesons and baryons in QCD in two dimensions \cite{Svetitsky1982, Yaffe1982, Delfino2, Delfino3, Delfino5}. 

\item \texttt{Potts}$_{4}$. The $4$-states Potts model can describe the deconfinement of baryons and mesons as in the case $q=3$ as well as tetraquark confined states which are allowed for $q=4$ only. Moreover the Ashkin-Teller model is a $\mathbb{Z}_{4}$-symmetric model which is represented by the following Hamiltonian
$$
\mathcal{H}_{\text{AT}}=-J\sum_{\langle ij\rangle}(\sigma_{i}\sigma_{j}+\tau_{i}\tau_{j})-J_{4}\sum_{\langle ij\rangle}(\sigma_{i}\sigma_{j}\tau_{i}\tau_{j})
$$
and since $\sigma$ and $\tau$ variables are equal to $\pm1$, but also
$\sigma\tau=\pm1$, when $J=J_{4}$ we can rephrase the problem to
have $4-1=3$ invariants. This particular point is explicitly $S_{4}$
symmetric and can be described by a $4$-states Potts model \cite{Delfino4}.
\end{itemize}
In the next section we are going to introduce the continuous formulation of the Potts model (Potts field theory) which describes the scaling limit of the RCM for $q\in\mathbb{R}$.

\subsection{Action and invariants}

As was shown by Golner \cite{Golner1973} and Zia and Wallace \cite{ZiaWallace1975}, the critical behavior of the $(n+1)$-state Potts model in $d$-dimensions can be studied by a $n$-component
bosonic field $\phi_{i}$ ($i=1,...,n$) carrying a representation of the $S_{n+1}$-symmetry. 
This representation involves the set of  the $n+1$ vectors $e^{\alpha}_i$  ($\alpha=1,...,n+1$) pointing to the vertices of the $n$-simplex, so that the underlying symmetry of the model, arising from the equivalence of its ($n+1$)-states, is reflected by the fact that the indices $\alpha,\beta,\gamma,\dots$ can be permuted amongst each other without changing the physics. 
The corresponding action is therefore symmetric under the discrete group which maps the $n$-dimensional hyper-tetrahedron on itself and this group isomorphic to $S_{n+1}$ \cite{Amit1976, Priest1976, DeAlcantara1980}.

%We now shortly classify the invariants and construct the action for an $n$-component scalar 
%%The fundamental difference between a $p$-dimensional vector and tensor
%order parameter is that the latter have more invariants with respect
%to the $p$-dimensional rotation group than the former. Thus both
%vectors and tensors have one quadratic invariant, but tensors have
%one cubic invariant instead of zero, and two quartic invariants instead
%of one
%Da citare: \cite{Priest1976}.

\subsubsection{Simplex}

In geometry a $simplex$  is a generalization of the notion
of a triangle or tetrahedron to arbitrary dimensions. Specifically, a regular $k$-simplex
is a $k$-dimensional polytope which is the convex hull of its $k+1$
vertices. For example, a $2$-simplex is a triangle, a $3$-simplex
is a tetrahedron, and so on (see Figure~\ref{sim}).
We can embed the regular $n$-dimensional simplex in $\mathbb{R}^{n}$ by writing directly its cartesian components.
This can be achieved with the help of the following two properties:
\begin{enumerate}
\item the distances of the vertices to the center are all equal
\item the angle subtended by any two vertices through its center is $\arccos(-\frac{1}{n})$\,.
\end{enumerate}
These properties allow the explicit construction of the vectors $e_{i}^{\alpha}$ that we will use in the definition of the $S_{n+1}$-invariants.
Explicit manipulations of expressions involving $e_{i}^{\alpha}$ can be performed using the following relations:
\begin{eqnarray}
\sum_{i=1}^n e_{i}^{\alpha}e_{i}^{\beta} & =&(n+1)\delta^{\alpha\beta}-1\label{rule_1}\nonumber\\
\sum_{\alpha=1}^{n+1} e_{i}^{\alpha} & =&0\label{rule_2}\nonumber\\
\sum_{\alpha=1}^{n+1} e_{i}^{\alpha}e_{j}^{\alpha} & =&(n+1)\delta_{ij}\,.\label{rule_3}
\end{eqnarray}
Note that in terms of the rules \eqref{rule_3}, the vectors $e^{\alpha}$ are normalised such that $e^{\alpha}\cdot e^{\alpha}=n$. This choice turns out to be useful in order to be able to take the limit $n\to0$ later on.
The rules \eqref{rule_3} are the basic relations that we will use in section \ref{FRG} to reduce the traces involved in the computation of the beta functions.

\subsubsection{Invariants}

\begin{figure}
\begin{centering}
\includegraphics[scale=0.3]{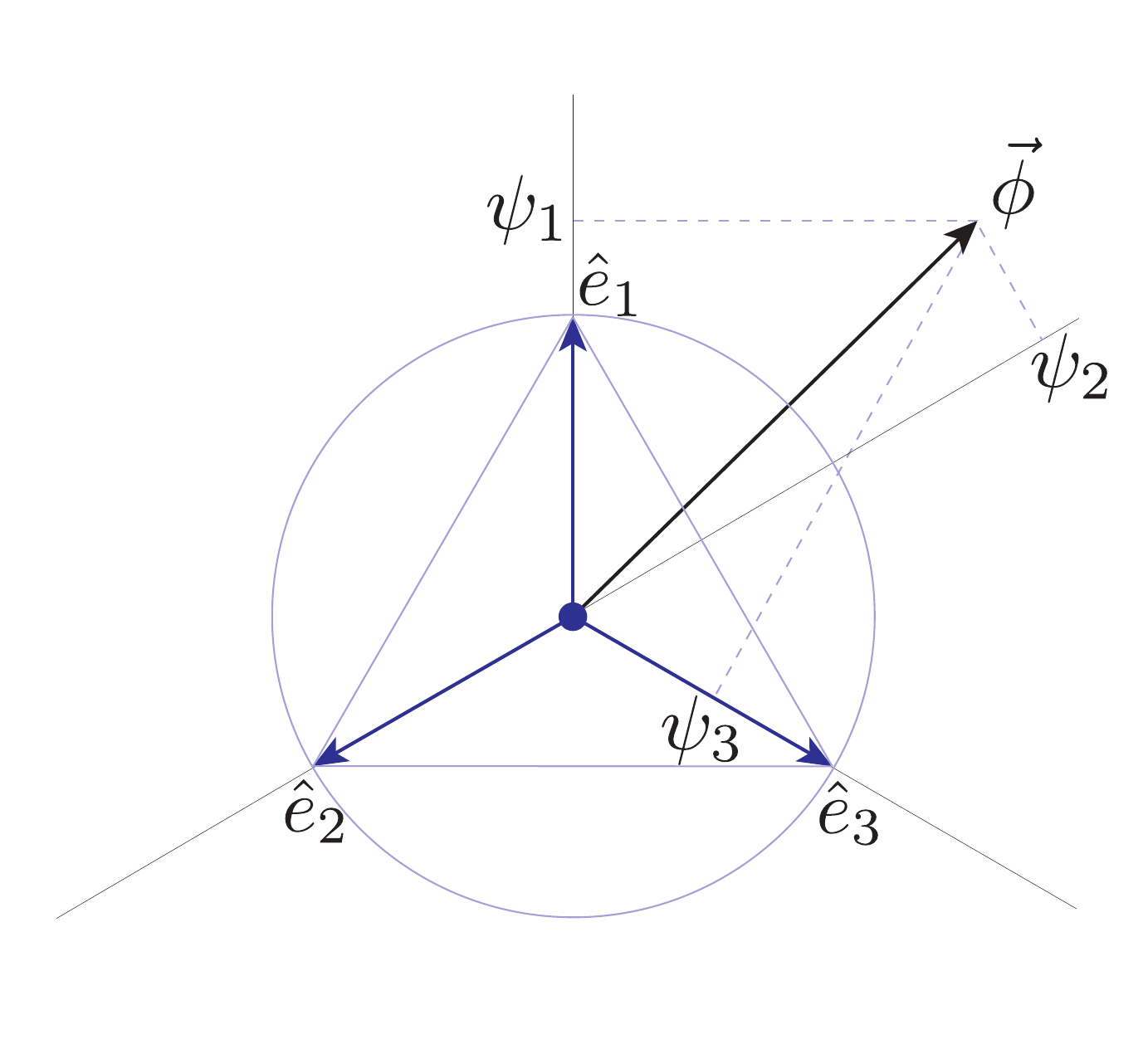}
\par\end{centering}
\caption{Relation between the fields $\phi_i$ and $\psi^\alpha$ in the $n=2$ case when $S_3$ is the symmetry group of an equilateral triangle in the plane.\label{fig2}}
\end{figure}

A simple way to understand how to construct invariants under the permutation
group is to approach the problem geometrically.
%An equilateral triangle is inscribed in the circle, a tetrahedron in the sphere and a $n$-hyper-tetrahedron in the $(n-1)$-dimensional sphere.
Consider a $n$-component scalar $\phi_{i}$ in $\mathbb{R}^{n}$ as our fluctuating field.
First we can construct the projections along the vectors of the hyper-tetrahedron
defining the ($n+1$)-fields $\psi^{\alpha}\equiv e{}^{\alpha}_{i}\cdot\phi_{i}$.
To aid the intuition consider a regular triangle in the plane as in Figure~\ref{fig2}.
%and more generally a regular polygon. In the limit this polygon becomes
%a circle then $e^{\alpha}_{i}\to\delta^{i}_{i}$ and the field $\psi^{\alpha}$
%and $\phi_{i}$ become equivalent.
%
Invariants are easy constructed in terms of the fields $\psi^{\alpha}$:
any symmetric polynomial will be invariant. In particular we consider
the \emph{power sum symmetric polynomials}, which are a type of basic
building block for symmetric polynomials, in the sense that every
symmetric polynomial with rational coefficients can be expressed as
a sum and difference of products of power sum symmetric polynomials
with rational coefficients. These are defined as
\begin{equation}
P_{k}=\sum_{\alpha=1}^{n+1}(\psi^{\alpha})^k\,.
\end{equation}
Note that $\sum_{\alpha}e_{i}^{\alpha}=0$ in \eqref{rule_3} implies $P_{1}=\sum_{\alpha}\psi^{\alpha}=0$,
because obviously not all the fields $\psi^{\alpha}$ are independent.
Since any symmetric polynomial in $\psi^{1},...,\psi^{n+1}$ can be
expressed as a polynomial expression with rational coefficients
%(additions and multiplications, possibly involving rational coefficients)
in the power sum symmetric polynomials, then it is evident that invariants
are monomials in the variables $\{P_{2},...,P_{n+1}\}$.
Once the field power $p$ has been fixed 
the number of invariants at each order is easily seen to be given
by $N(p)=P(p)-P(p-1)$, where $P(p)$ is the partition of $p$ objects.
Starting from $p=2$, the number of invariants is then given by the
sequence $1,1,2,2,4,4,7,8,12,14,\dots$ More specifically,
for $p=1$ and $p=2$ we have $P_{2}$ and $P_{3}$; for $p=4$ we
have $P_{4}$ and $P_{2}^{2}$; for $p=5$ the invariants are $P_{5}$
and $P_{2}P_{3}$; while for $p=6$ there are four possibilities:
$P_{6}$, $P_{3}^{2}$, $P_{2}P_{4}$ and $P_{2}^{3}$; and so on.

The basic invariants $P_{k}$ can be expressed back in terms of the
fields $\phi_{i}$. For example 
\begin{equation*}
P_{2}=\sum_{\alpha}(\psi^{\alpha})^2=\sum_{\alpha}e^{\alpha}_{i}e^{\alpha}_{j}\,\phi_{i}\phi_{j}=(n+1)\delta_{ij}\,\phi_{i}\phi_{j}\,,
\end{equation*}
where we used the last rules in (\ref{rule_3}). Similarly
\begin{equation*}
P_{3}=\sum_{\alpha}(\psi^{\alpha})^3=\sum_{\alpha}e^{\alpha}_{i}e^{\alpha}_{j}e^{\alpha}_{k}\,\phi_{i}\phi_{j}\phi_{k}\,,
\end{equation*}
and more generally
\begin{equation}
P_{k}=\sum_{\alpha}e_{i_{1}}^{\alpha}\dots e_{i_{k}}^{\alpha}\,\phi_{i_{1}}\cdots\phi_{i_{k}}\,.
\end{equation}
At this point it is convenient to define the tensors 
\begin{equation}\label{eq:tensordef1}
T_{i_{1}\dots i_{p}}^{(p,1)}\equiv\sum_{\alpha}e_{i_{1}}^{\alpha}\dots e_{i_{p}}^{\alpha}\qquad\qquad p\geq3\,,
\end{equation}
with the exception $T_{i_{1}i_{2}}^{(2,1)}\equiv\delta_{i_{1}i_{2}}$
when $p=2$ (we factor away an inessential $n+1$). When an invariant
is the product of two or more power polynomials it can be reduced
to a product of the tensors just defined. For example
\begin{eqnarray*}
\ensuremath{P_{2}P_{4}} & = & \sum_{\alpha}(\psi^{\alpha})^2\sum_{\beta}(\psi^{\beta})^4
%\\ & = & 
=\sum_{\alpha}e_{i_{1}}^{\alpha}e_{i_{2}}^{\alpha}\sum_{\beta}e_{i_{3}}^{\beta}e_{i_{4}}^{\beta}e_{i_{5}}^{\beta}e_{i_{6}}^{\beta}\,\phi_{i_{1}}\cdots\phi_{i_{6}}\\
 & = & (n+1)\delta_{i_{1}i_{2}}\sum_{\beta}e_{i_{3}}^{\beta}e_{i_{4}}^{\beta}e_{i_{5}}^{\beta}e_{i_{6}}^{\beta}\,\phi_{i_{1}}\cdots\phi_{i_{6}}
 %\\& = & 
 =(n+1)\left\{ T_{i_{1}i_{2}}^{(2,1)}T_{i_{2}i_{3}i_{4}i_{5}}^{(4,1)}\right\} \phi_{i_{1}}\cdots\phi_{i_{6}}\,,
\end{eqnarray*}
or
\begin{eqnarray*}
\ensuremath{P_{3}^{2}&=&\sum_{\alpha}(\psi^{\alpha})^{3}\sum_{\beta}(\psi^{\beta})^{3}=\left(\sum_{\alpha}e_{i_{1}}^{\alpha}e_{i_{2}}^{\alpha}e_{i_{3}}^{\alpha}\right)\left(\sum_{\beta}e_{i_{4}}^{\beta}e_{i_{5}}^{\beta}e_{i_{6}}^{\beta}\right)\phi_{i_{1}}\cdots\phi_{i_{6}}\\
&=&\left\{ T_{i_{1}i_{2}i_{3}}^{(3,1)}T_{i_{4}i_{5}i_{6}}^{(3,1)}\right\} \phi_{i_{1}}\cdots\phi_{i_{6}}}\,,
\end{eqnarray*}
and similarly for all other possible cases. We remark here that a symmetrization over all indexes is intended when needed. Thus at any order $p$ we can define
$N(p)$ tensors $T_{i_{1}\dots i_{p}}^{(p,m)}$ with $m=1,...,N(p)$
as shown in Table \ref{T_inv},
%, constructed from the building blocks $T_{i_{1}\dots i_{p}}^{(p,1)}$,
%which, 
that when contracted with $p$-fields $\phi_{i}$ constitute a basis for non-derivative invariants.% of order $\phi^p$.
%of invariants.
%
\begin{table}
\begin{centering}
\begin{tabular}{c|c|c|c|c|c}
$T^{(p,m)}$ & $m=1$ & $m=2$ & $m=3$ & $m=4$ & $N(p)$ \tabularnewline
\hline 
\hline 
$p=2$ & $\delta$  &  &  &  & $1$\tabularnewline
\hline 
$p=3$ & $\sum eee$  &  &  &  & $1$\tabularnewline
\hline 
$p=4$ & $\sum eeee$ & $\delta\delta$ &  &  & $2$\tabularnewline
\hline 
$p=5$ & $\sum eeeee$ & $\delta\sum eee$ &  &  & $2$\tabularnewline
\hline 
$p=6$ & $\sum eeeeee$ & $\sum eee\sum eee$ & $\delta\sum eeee$ & $\delta\delta\delta$ & $4$\tabularnewline
\end{tabular}
\par\end{centering}
\caption{Tensor invariants $T^{(p,m)}$ defined in the text with their relative number $N(p)$. The index structure of the invariants follows the general structure of equation \eqref{eq:tensordef1}.  \label{T_inv}}
\end{table}

\subsubsection{Action}

We can now construct the general action invariant under the permutation group containing non-derivative interactions.
Introducing the convenient notation $T_{i_{1}\dots i_{p}}^{(p)}$ for the tensor coupling of $p$-fields we can compactly write
\begin{eqnarray}
S[\phi] & = & \int_{x}\left\{ \frac{1}{2}\partial^{\mu}\phi_{i}\partial_{\mu}\phi_{i}+\frac{1}{2}T_{i_{1}i_{2}}^{(2)}\phi_{i_{1}}\phi_{i_{2}}+\frac{1}{3!}T_{i_{1}i_{2}i_{3}}^{(3)}\phi_{i_{1}}\phi_{i_{2}}\phi_{i_{3}}+\right.\label{eq:acgenexp}\nonumber\\
 &  & \left.+\frac{1}{4!}T_{i_{1}i_{2}i_{3}i_{4}}^{(4)}\phi_{i_{1}}\phi_{i_{2}}\phi_{i_{3}}\phi_{i_{4}}+\frac{1}{5!}T_{i_{1}i_{2}i_{3}i_{4}i_{5}}^{(5)}\phi_{i_{1}}\phi_{i_{2}}\phi_{i_{3}}\phi_{i_{4}}\phi_{i_{5}}+\dots\right\} \,, 
\end{eqnarray}
which implicitly defines the dimensionful couplings $\bar{\lambda}_{p,m}$ (clearly one for each invariant of Table \ref{T_inv}), 
\begin{eqnarray}
&&T_{i_{1}i_{2}i_{3}}^{(3)}  =  \bar{\lambda}_{3}T_{i_{1}i_{2}i_{3}}^{(3,1)}\nonumber\\
&&T_{i_{1}i_{2}i_{3}i_{4}}^{(4)}  =  \bar{\lambda}_{4,1}T_{i_{1}i_{2}i_{3}i_{4}}^{(4,1)}+\bar{\lambda}_{4,2}T_{i_{1}i_{2}i_{3}i_{4}}^{(4,2)}\nonumber\\
&&T_{i_{1}i_{2}i_{3}i_{4}i_{5}}^{(5)}  =  \bar{\lambda}_{5,1}T_{i_{1}i_{2}i_{3}i_{4}i_{5}}^{(5,1)}+\bar{\lambda}_{5,2}T_{i_{1}i_{2}i_{3}i_{4}i_{5}}^{(5,2)}\nonumber\\
&&T_{i_{1}i_{2}i_{3}i_{4}i_{5}i_{6}}^{(6)}  =  \bar\lambda_{6,1}T_{i_{1}i_{2}i_{3}i_{4}i_{5}i_{6}}^{(6,1)}+\bar\lambda_{6,2}T_{i_{1}i_{2}i_{3}i_{4}i_{5}i_{6}}^{(6,2)}
%\\ &  &
 +\bar\lambda_{6,3}T_{i_{1}i_{2}i_{3}i_{4}i_{5}i_{6}}^{(6,3)}+\bar\lambda_{6,4}T_{i_{1}i_{2}i_{3}i_{4}i_{5}i_{6}}^{(6,4)}\,.
\end{eqnarray}
Finally, we define the following invariants through contraction%CORRECT
\begin{table}\label{tab:inv}
\begin{centering}
\begin{tabular}{c||ccc|ccc|ccc}
 &  & $n=1$ &  &  & $n=2$ &  &  & $n=3$ & \tabularnewline
\hline 
$I_{2}$ &  & $\varphi_{1}^{2}$ &  &  & $\varphi_{1}^{2}+\varphi_{2}^{2}$ &  &  & $\varphi_{1}^{2}+\varphi_{2}^{2}+\varphi_{3}^{2}$ & \tabularnewline
\hline 
$I_{3}$ &  & $0$ &  &  & $\frac{3}{\sqrt{2}}\varphi_{2}(\varphi_{2}^{2}-3\varphi_{1}^{2})$ &  &  & $\frac{4}{\sqrt{3}}\left(\sqrt{2}\varphi_{1}(\varphi_{1}^{2}-3\varphi_{2}^{2})-3(\varphi_{1}^{2}+\varphi_{2}^{2})\varphi_{3}+2\varphi_{3}^{3}\right)$ & \tabularnewline
\hline 
$I_{4,1}$ &  & $2I_{2}^{2}$ &  &  & $\frac{9}{2}I_{2}^{2}$ &  &  & \texttt{\small{}$\begin{array}{c}
8\!\left(\varphi_{1}^{4}+\varphi_{2}^{4}+\frac{7}{6}\varphi_{3}^{4}+\varphi_{1}^{2}(\varphi_{3}^{2}-\frac{2\sqrt{2}}{3}\varphi_{1}\varphi_{3})\right.\qquad\quad\\
\qquad\qquad\qquad\quad\left.+2\varphi_{2}^{2}(\varphi_{1}^{2}+\sqrt{2}\varphi_{1}\varphi_{3}+\frac{1}{2}\varphi_{3}^{2})\right)
\end{array}$} & \tabularnewline
\hline 
$I_{4,2}$ &  & $I_{2}^{2}$ &  &  & $I_{2}^{2}$ &  &  & $I_{2}^{2}$ & \tabularnewline
\hline 
$I_{5,1}$ &  & $0$ &  &  & $\frac{5}{2}I_{2}I_{3}$ &  &  & $\frac{10}{3}I_{2}I_{3}$ & \tabularnewline
\hline 
$I_{5,2}$ &  & $0$ &  &  & $I_{2}I_{3}$ &  &  & $I_{2}I_{3}$ & \tabularnewline
\hline 
$I_{6,1}$ &  & $2I_{2}^{\text{3}}$ &  &  & $\frac{27}{4}I_{2}^{3}+I_{3}^{2}$ &  &  & $\frac{1}{3}I_{3}^{2}+3I_{2}I_{4,1}-8I_{2}^{3}$ & \tabularnewline
\hline 
$I_{6,2}$ &  & $2I_{2}^{\text{3}}$ &  &  & $\frac{9}{2}I_{2}^{3}$ &  &  & $I_{2}I_{4,1}$ & \tabularnewline
\hline 
$I_{6,3}$ &  & $I_{2}^{\text{3}}$ &  &  & $I_{2}^{3}$ &  &  & $I_{2}^{3}$ & \tabularnewline
\hline 
$I_{6,4}$ &  & $0$ &  &  & $I_{3}^{2}$ &  &  & $I_{\text{3}}^{2}$ & \tabularnewline
\end{tabular}
\par\end{centering}
\caption{Explicit form of the invariants of $S_{n+1}$ for $n=1,2,3$. 
The first $n$ invariants constitute a basis with rational
coefficients upon which all other invariants defined in \eqref{perminv} can be expressed. \label{T_inv2}}
\end{table}
\begin{equation}
I_{p,m}\equiv T_{i_{1}\cdots i_{p}}^{(p,m)}\phi_{i_{1}}\cdots\phi_{i_p}\,,
\label{perminv}
\end{equation}
which are clearly related to the invariants $P_{k}$ of the preceding subsection, but now function of the fields $\phi_i$,
and re-write the action \eqref{eq:acgenexp} as
\begin{equation}
S[\phi]=\int_{x}\left\{ \frac{1}{2}\partial^{\mu}\phi_{i}\partial_{\mu}\phi_{i}+V(\phi_1,..,\phi_n)\right\}\,, 
\label{eq:b_action}
\end{equation}
where the potential is constructed from the invariants \eqref{perminv} as follows% of the permutation group
% a function of the $n$ independent invariants of $S_{n+1}$
%
\begin{equation}\label{eq:Vexp}
V(\phi_1,..,\phi_n) 
 %& = & \frac{1}{2}\bar{\lambda}_{2}I_{2}+\frac{1}{3!}\bar{\lambda}_{3}I_{3}+\frac{1}{4!}(\bar{\lambda}_{4,1}I_{4,1}+\bar{\lambda}_{4,2}I_{4,2})
%+\frac{1}{5!}(\bar{\lambda}_{5,1}I_{5,1}+\bar{\lambda}_{5,2}I_{5,2})+\dots\notag\\
  =  \sum_{p=2}^{\infty}\frac{1}{p!}\sum_{m=1}^{N(p)}\bar{\lambda}_{p,m}I_{p,m}
  = \frac{1}{2}\bar{\lambda}_{2}I_{2}+\frac{1}{3!}\bar{\lambda}_{3}I_{3}+\frac{1}{4!}(\bar{\lambda}_{4,1}I_{4,1}+\bar{\lambda}_{4,2}I_{4,2})+...%\notag
\end{equation}
But, as expected, only the first $n$ invariants are independent, %under $S_{n+1}$
as can be seen from the explicitly construct reported in Table \ref{T_inv2} in the cases $n=1,2,3$.
%and the potential \eqref{eq:Vexp} is thus function only of these. 
The examples of Table \ref{T_inv2}  also show that  indeed rational coefficients are
needed to express the dependent invariants in terms of the independent ones.

\subsection{Explicit construction for $n=2$}
To facilitate the understanding we give an explicit construction of the Potts field theory in the $S_3$ case.
The coordinates of the vertices of a regular triangle in the plane, as shown in Figure \ref{fig2}, are
{\small
\begin{equation}
e^{1}=\sqrt{2}\left(\begin{array}{c}
0\\
1
\end{array}\right)\qquad e^{2}=\sqrt{2}\left(\begin{array}{c}
-\frac{\sqrt{3}}{2}\\
-\frac{1}{2}
\end{array}\right)\qquad e^{3}=\sqrt{2}\left(\begin{array}{c}
\frac{\sqrt{3}}{2}\\
-\frac{1}{2}
\end{array}\right)\,.
\end{equation}
}
\!\!The matrix representation of the $|S_3|=3!=6$ elements of $S_{3}$, which leave
invariant the triangle, are
{\small
\begin{align}
I & =\left(\begin{array}{cc}
1 & 0\\
0 & 1
\end{array}\right)\qquad R=\left(\begin{array}{cc}
-\frac{1}{2} & -\frac{\sqrt{3}}{2}\\
+\frac{\sqrt{3}}{2} & -\frac{1}{2}
\end{array}\right)\qquad R^{-1}=\left(\begin{array}{cc}
-\frac{1}{2} & \frac{\sqrt{3}}{2}\\
-\frac{\sqrt{3}}{2} & -\frac{1}{2}
\end{array}\right)\nonumber\\
\mu_1 & =\left(\begin{array}{cc}
-1 & 0\\
0 & 1
\end{array}\right)
\qquad \mu_2=\left(\begin{array}{cc}
+\frac{1}{2} & +\frac{\sqrt{3}}{2}\\
+\frac{\sqrt{3}}{2} & -\frac{1}{2}
\end{array}\right)
\qquad \mu_3=\left(\begin{array}{cc}
+\frac{1}{2} & -\frac{\sqrt{3}}{2}\\
-\frac{\sqrt{3}}{2} & -\frac{1}{2}
\end{array}\right)\,.
\label{trans}
\end{align}
}
\!\!Clearly $I$ is the identity, $R$ the (counter-clockwise) rotation of $\frac{2}{3}\pi$ and $R^{-1}$ its inverse, while $\mu_i$ for $i=1,2,3$ are the reflections along the axis passing through the $i$-th vertex of the triangle.
The two $n=2$ invariants are, from Table \ref{T_inv2}, the following
$$
\rho\equiv I_{2}=3(\varphi_{1}^{2}+\varphi_{2}^{2})\qquad\qquad\tau\equiv I_{3}=\frac{3}{\sqrt{2}}\varphi_{2}(\varphi_{2}^{2}-3\varphi_{1}^{2})\,.
$$
We can check explicitly that these invariants are indeed so under the
transformations \eqref{trans}. Consider for example the rotation $R
$\begin{equation*}
\left(\begin{array}{c}
\varphi_{1}\\
\varphi_{2}
\end{array}\right)\longmapsto\left(\begin{array}{c}
\tilde{\varphi}_{1}\\
\tilde{\varphi}_{2}
\end{array}\right)=R\left(\begin{array}{c}
\varphi_{1}\\
\varphi_{2}
\end{array}\right)=\left(\begin{array}{c}
-\frac{1}{2}\varphi_{1}-\frac{\sqrt{3}}{2}\varphi_{2}\\
\frac{\sqrt{3}}{2}\varphi_{1}-\frac{1}{2}\varphi_{2}
\end{array}\right)\,.
\end{equation*}
Its easy to check the invariance of $\rho$ and $\tau$
{\small
\begin{align*}
\tilde{\rho} & =3(\tilde{\varphi}_{1}^{2}+\tilde{\varphi}_{2}^{2})\\
 & =\frac{3}{4}\varphi_{1}^{2}+\frac{9}{4}\varphi_{2}^{2}+\frac{3\sqrt{3}}{2}\varphi_{1}\varphi_{2}+\frac{3}{4}\varphi_{2}^{2}+\frac{9}{4}\varphi_{1}^{2}-\frac{3\sqrt{3}}{2}\varphi_{1}\varphi_{2}\\
 & =3(\varphi_{1}^{2}+\varphi_{2}^{2})=\rho
\end{align*}
\begin{align*}
\tilde{\tau} & =\frac{3}{\sqrt{2}}\tilde{\varphi}_{2}(\tilde{\varphi}_{2}^{2}-3\tilde{\varphi}_{1}^{2})\\
 & =\frac{3}{\sqrt{2}}\left(\frac{\sqrt{3}}{2}\varphi_{1}-\frac{1}{2}\varphi_{2}\right)\left[\left(\frac{\sqrt{3}}{2}\varphi_{1}-\frac{1}{2}\varphi_{2}\right)^{2}-3\left(-\frac{1}{2}\varphi_{1}-\frac{\sqrt{3}}{2}\varphi_{2}\right)^{2}\right]\\
 & =\frac{3}{\sqrt{2}}\varphi_{2}(\varphi_{2}^{2}-3\varphi_{1}^{2})=\tau\,.
\end{align*}
}
\!Similarly one can check all the other transformations in \eqref{trans}. So we showed explicitly that
the theory with potential of the form \eqref{eq:Vexp}
\begin{equation}
V(\varphi_{1},\varphi_{2})=\frac{\bar{\lambda}_{2}}{2}3\left(\varphi_{1}^{2}+\varphi_{2}^{2}\right)+\frac{\bar{\lambda}_{3}}{3!}\frac{3}{\sqrt{2}}\varphi_{2}(\varphi_{2}^{2}-3\varphi_{1}^{2})+...
\end{equation}
is invariant under the action of the elements of $S_{3}$.

\section{Functional RG for Potts\label{FRG}}

\subsection{Flow equation}

The functional renormalization group (FRG) approach to quantum field theory  is based on the exact flow equation satisfied by the scale dependent effective action $\Gamma_{k}$ (for a general review see e.g. \cite{Wett2}; while for a self-contained introduction with particular attention to statistical physics see \cite{delamotte1}). This is a scale-dependent functional which includes fluctuations between a given microscopic UV scale $\Lambda$ down to a running scale $k<\Lambda$. The effective action  interpolates smoothly between the bare UV action $S =\Gamma_{k=\Lambda} $ and the full effective action, or free energy, $\Gamma =\Gamma_{k=0} $ for $k\to0$ so that all fluctuations are summed over.
The scale dependence of the effective action $\Gamma_{k}$ 
on the RG time $t:=\log k$ is governed by the exact flow equation \cite{Wett1}, which for an $n$-component scalar $\varphi_i  \equiv \left\langle \phi_i \right\rangle $ reads
\begin{equation}\label{eq:Wett}
\partial_{t}\Gamma_{k}[\varphi]=\frac{1}{2}{\rm Tr}\left(\frac{\delta^{2}\Gamma_{k}[\varphi]}{\delta\varphi_{i}\delta\varphi_{j}}+R_{k,ij}\right)^{-1}\partial_{t}R_{k,ji}\,.
 \end{equation}
Here $R_k$ is a proper infrared regulator function which suppresses the propagation of the infrared modes (of momentum smaller than $k$) by directly modifying the bare propagator of the theory.
The Wetterich equation \eqref{eq:Wett} is the starting point of all our subsequent analysis.
%In the following we will use the flow equaiton \eqref{eq:Wett} to extract the beta functions 
%Several cutoff function have been proposed in the literature and though implementing different coarse graining, they all ensure the aforementioned constraints that $\Gamma_k$ approaches the bare action in the UV limit  $(k\to\Lambda)$  and  the  full  quantum  effective action  in the IR limit $(k\to 0)$.
%
\subsubsection{Local potential approximation}
Despite its simplicity, the Wetterich equation \eqref{eq:Wett} is difficult to solve and one should rely on approximations based  on  non-perturbative  truncations, which amounts to project the RG flow on a subset of suitable functionals. One of these truncations is called improved local potential approximation (LPA') and consist in considering the following ansatz for the effective-action %({\it Usare $I_{a,b}$?})
\begin{equation}\label{eq:ansatz}
\Gamma_{k}[\varphi]=\int_{x}\left\{ \frac{1}{2}Z_{k}\partial^{\mu}\varphi_{i}\partial_{\mu}\varphi_{i}+V_{k}(\varphi_1,...,\varphi_n)\right\}\,,
\end{equation}
where the whole theory space is projected into the infinite dimensional functional space of effective potentials $V_k$. At first order of the  derivative expansion, also called local potential approximation (LPA), one neglects the running and the field dependence of the wave function renormalization, $Z_k\equiv 1$. In the improved local potential approximation (LPA') we consider throughout, $Z_k$ is a  non-vanishing field-independent but scale-dependent running wavefunction renormalization constant, directly related to the anomalous dimension $\eta_k = -\partial_t \log Z_k$. 

We can obtain a flow equation for the effective potential by inserting the ansatz \eqref{eq:ansatz} in the Wetterich equation \eqref{eq:Wett} .
The first thing to do is to compute the Hessian which, dropping for the moment $Z_k$, reads
\begin{equation}\label{hessian}
\frac{\delta^{2}\Gamma_{k}}{\delta\varphi_{i}\delta\varphi_{j}}=-\partial^{2}\delta_{ij}+V_{k,ij}=:-\partial^{2}(\mathbb{I})_{ij}+(\mathbb{V})_{ij}\,,
\end{equation}
%
%$$
%(\mathbb{V}_{k}){}_{ij}=\frac{\partial^{2}V_{k}}{\partial\varphi_{i}\partial\varphi_{j}}\,.
%$$
%
where we introduced matrix notation for clarity. When we insert \eqref{hessian} in
the flow equation \eqref{eq:Wett}, with the choice $R_{k,ij} = \delta_{ij} R_k$ for the matrix structure of the cutoff, we find 
\begin{equation}
\partial_{t}\Gamma_{k}=\frac{1}{2}{\rm Tr}\frac{\partial_{t}R_{k}(-\partial^{2})}{(-\partial^{2}+R_{k}(-\partial^{2}))\mathbb{I}+\mathbb{V}}\,.
\end{equation}
After choosing constant field configuration so that $\partial_{t}\Gamma_{k}= (\int \rm{d}^d x)\,\partial_{t}V_{k}$
and performing the angular integrations, we obtain the following expression
for the flow of the potential 
\begin{equation}\label{dVgen}
\partial_{t}V_{k}=\frac{1}{2}\frac{1}{(4\pi)^{\frac{d}{2}}\Gamma(\frac{d}{2})}\int_{0}^{\infty}{\rm d}z\,z^{\frac{d}{2}-1}\,{\rm tr}\,\frac{\partial_{t}R_{k}(z)}{\left(z+R_{k}(z)\right)\mathbb{I}+\mathbb{V}}\,.
\end{equation}
We will adopt now the linear cutoff $R_{k}(z)=(k^{2}-z)\theta(k^{2}-z)$, where $\theta$ is the standard Heaviside step function, that allows a simple explicit evaluation of the integral in \eqref{dVgen}.
%This cut-off function is chosen to optimize the results of the truncation of equation\eqref{eq:ansatz} when scalar field theories are considered \cite{Litim1, Litim2}. 
With this choice we find the following form for the flow equation of the potential
\begin{equation}\label{eq:flow}
\partial_{t}V_{k}=c_{d} k^{d+2}\,{\rm tr}\frac{1}{k^{2}\mathbb{I}+\mathbb{V}}\,,
\end{equation}
where we defined the constant $c_{d}^{-1}\equiv(4\pi)^{\frac{d}{2}}\Gamma(\frac{d}{2}+1)$.  This expression is the general form for the LPA of an $n$-component scalar in $d$-dimensions and as such it is the generating function of all beta functions of the couplings $\bar \lambda_{p,m}$ of non-derivative interactions entering \eqref{eq:Vexp}.
%
%This equation is valid in any dimension and for any value of $n$.% and is the 

Unfortunately,  for general $n$, it is not possible to obtain a closed form for the inverse of the matrix $k^{2}\mathbb{I}+\mathbb{V}$.
Therefore, since our main interest is the study of the limits $n\to 0$ ({\tt Percolation}) and  $n\to -1$ ({\tt Spanning Forest}) (which we remember are the only non-trivial cases apart {\tt Ising} in $d>2$) we are forced to {\it truncate} the effective potential \eqref{eq:Vexp} in order to convert  \eqref{eq:flow} in a set of coupled beta functions (explicitly $n$-dependent) for a {\it finite} set of couplings, by expanding the r.h.s. of \eqref{eq:flow} in powers of $\mathbb{V}$ and thus reducing the problem to the evaluation of traces of these powers. This approach is presented in section \ref{sec:betan}, where the appropriate "trace machinery" will be developed and where we report the truncation up to $\varphi^6$.  A different approach is  based on the fact that the inversion of the matrix in \eqref{eq:flow} is instead possible whenever $n$ is a given, conceivably small, positive integer (and thus not applicable in the {\tt Percolation} and {\tt Spanning Forest} cases) and in this case we are able to write the explicit form of the LPA'.  This is presented in section \ref{LPA} for the $n=1,2$ cases. 
\subsection{Beta functions for general $n$}\label{sec:betan}
The aim of this section is to provide a general framework to extract beta functions for any value of $n\in \mathbb{R}$. Defining $\mathbb{V}=\bar \lambda_{2}\mathbb{I}+\mathbb{M}$ we proceed expanding the inverse propagator as
\begin{equation}\label{trMM}
{\rm tr}\frac{1}{k^{2}\mathbb{I}+\mathbb{V}}={\rm tr}\,\frac{k^{-2}}{(1+\lambda_{2})\mathbb{I}+\mathbb{M}/k^2}
=\sum_{m=0}^{\infty}(-1)^{m}\frac{k^{-2-2m}}{(1+\lambda_{2})^{m+1}}\text{tr}\,\mathbb{M}^{m}\,,
\end{equation}
where from now on we absorb the factor $c_d$ in the definition of the effective potential and of the field $V_k\to c_d V_k$, $\varphi_i\to c_d^{1/2}\varphi_i$. Inserting \eqref{trMM} into the  flow equation of the effective potential  \eqref{eq:flow}  gives
\begin{equation}\label{eq:flow2}
\partial_{t}V_{k}=\sum_{m=0}^{\infty} (-1)^{m}\frac{k^{d-2m}}{(1+\lambda_{2})^{m+1}}\text{tr}\,\mathbb{M}^{m}\,.
\end{equation}
Any field dependence on the r.h.s. of equation \eqref{eq:flow2} is encoded in $\text{tr}\,\mathbb{M}^m$ since
\begin{equation}
(\mathbb{M})_{ab}=T_{abi_{1}}^{(3)}\varphi_{i_{1}}+\frac{1}{2}T_{abi_{1}i_{2}}^{(4)}\varphi_{i_{1}}\varphi_{i_{2}}+\frac{1}{3!}T_{abi_{1}i_{2}i_{3}}^{(5)}\varphi_{i_{1}}\varphi_{i_{2}}\varphi_{i_{3}}+\dots
\end{equation}
and once the traces $\text{tr}\,\mathbb{M}^{m}$ are computed with the help of rules \eqref{rule_3} and expressed in terms of the invariants \eqref{perminv}, the r.h.s. of the flow equation \eqref{eq:flow} assumes the form
\begin{equation}\label{eq:betaflow}
\partial_t V_k =\frac{1}{2}\bar{\beta}_{2}I_{2}+\frac{1}{3!}\bar{\beta}_{3}I_{3}+\frac{1}{4!}(\bar{\beta}_{4,1}I_{4,1}+\bar{\beta}_{4,2}I_{4,2})
 +\frac{1}{5!}(\bar{\beta}_{5,1}I_{5,1}+\bar{\beta}_{5,2}I_{5,2})+\dots\,,
\end{equation}
from which we can extract the beta functions of all the couplings included in the truncation of the potential \eqref{eq:Vexp}.
%
%Though in a perturbative analysis around the upper critical dimension it is sufficient to consider only the first relevant operator in the potential $V_k$ \cite{ZiaWallace1975}, namely $\varphi^3$ when $d_c=6$  and to $\varphi^4$ when $d_c=4$, in the FRG approach we shall consider the expansion \eqref{eq:Vexp} up to $\varphi^p$ with $p$ as large as possible. Differently from other symmetry groups where the number of invariants is constant with $p$, the case of $S_{n+1}$-symmetry is peculiar since the number of invariants at order $p$ is growing as $N(p)$, and this is the main reason why large truncations become difficult. 
%
%in powers of the fields since further contributions will be of higher orders and should not be considered.
%
Before presenting our results to order $\varphi^6$, in the following subsection we are going to work out the simple $\varphi^3$ example to see how this general procedure works at hand.

\subsubsection{Example: $\varphi^{3}$}\label{ssec:phi3}

When we truncate the expansion \eqref{eq:Vexp} at order $p=3$, only the trilinear coupling and the mass terms are present in the corresponding effective action $\Gamma_k$. Accordingly, the non-diagonal part of the Hessian is just $\mathbb{M}=\bar{\lambda}_{3}T_{abi}^{(3)}\varphi_{i}$ and therefore we have to consider the contributions $\text{tr}\,\mathbb{M}$, $\text{tr}\,\mathbb{M}^2$, $\text{tr}\,\mathbb{M}^3$ since $\text{tr}\,\mathbb{M}^m\sim\varphi^{i\geq m}$. To show explicitly how the computation of these traces works  we use a colour code for the rules \eqref{rule_3} in order to highlight when they play a role in the evaluation
%
%\begin{eqnarray*}
%{\color{red}\sum_{\alpha}e_{i}^{\alpha}} & = & 0\\
%{\color{cyellow}e_{i}^{\alpha}e_{i}^{\beta}} & = & (n+1)\delta^{\alpha\beta}-1\\
%{\color{cgreen}e_{i}^{\alpha}e_{j}^{\alpha}} & = & (n+1)\delta_{ij}\,.
%\end{eqnarray*}
%
%
\begin{equation}
{\color{red}\sum_{\alpha}e_{i}^{\alpha}}  =  0\qquad\qquad
{\color{cyellow}e_{i}^{\alpha}e_{i}^{\beta}}  =  (n+1)\delta^{\alpha\beta}-1\qquad\qquad
{\color{cgreen}e_{i}^{\alpha}e_{j}^{\alpha}}  =  (n+1)\delta_{ij}\,.
\end{equation}
Apart the linear trace which is zero,
\begin{equation}
\text{tr}\,\mathbb{M}=\bar{\lambda}_3\,T^{(3)}_{aai}\,\varphi_i=\bar{\lambda}_3\sum_{\alpha}{\color{cyellow}e^{\alpha}_a e^{\alpha}_a}e^{\alpha}_i\,\varphi_i=n\,\bar{\lambda}_3\,\sum_{\alpha} {\color{red}e^{\alpha}_i}=0\,,
\end{equation}
non-trivial contributions come from the trace of the square 
\begin{eqnarray}\label{eq:trm2}
\text{tr}\,\mathbf{\mathbb{M}}^{2} & = & T_{aji_{1}}^{(3,1)}\varphi_{i_{1}}T_{jai_{2}}^{(3,1)}\varphi_{i_{2}}\notag\\
 & = & \bar{\lambda}_{3}^{2}\left[\sum_{\alpha\beta}{\color{cyellow}e_{a}^{\alpha}}e_{j}^{\alpha}e_{i_{1}}^{\alpha}e_{j}^{\beta}{\color{cyellow}e_{a}^{\beta}}e_{i_{2}}^{\beta}\right]\varphi_{i_{1}}\varphi_{i_{2}}\notag\\
 & = & \bar{\lambda}_{3}^{2}\left[(n+1)\sum_{\alpha}{\color{cyellow}e_{j}^{\alpha}}{\color{cgreen}e_{i_{1}}^{\alpha}}{\color{cyellow}e_{j}^{\alpha}}{\color{cgreen}e_{i_{2}}^{\alpha}}-\sum_{\alpha\beta}{\color{cyellow}e_{j}^{\alpha}}e_{i_{1}}^{\alpha}{\color{cyellow}e_{j}^{\beta}}e_{i_{2}}^{\beta}\right]\varphi_{i_{1}}\varphi_{i_{2}}\notag\\
 & = & \bar{\lambda}_{3}^{2}\left[n(n+1)^{2}\delta_{i_{1}i_{2}}-(n+1)\sum_{\alpha}{\color{cgreen}e_{i_{1}}^{\alpha}e_{i_{2}}^{\alpha}}-\sum_{\alpha}{\color{red}e_{i_1}^{\alpha}}\sum_{\beta}{\color{red}e_{i_2}^{\beta}}\right]\varphi_{i_{1}}\varphi_{i_{2}}\notag\\
 & = & \bar{\lambda}_{3}^{2}(n+1)^{2}(n-1)I_2
\end{eqnarray}
and  from the trace of the cube 
%
%{\small
\begin{eqnarray}\label{trMMM}
\text{tr}\,\ensuremath{\mathbb{M}}^{3} & = & T_{abi_{1}}^{(3,1)}\varphi_{i_{1}}T_{bci_{2}}^{(3,1)}\varphi_{i_{2}}T_{cai_{3}}^{(3,1)}\varphi_{i_{3}}\notag\\
 & = & \bar{\lambda}_{3}^{3}\sum_{\alpha\beta\gamma}{\color{red}{\color{cyellow}e_{a}^{\alpha}}}e_{b}^{\alpha}e_{i_{1}}^{\alpha}e_{b}^{\beta}e_{c}^{\beta}e_{i_{2}}^{\beta}e_{c}^{\gamma}{\color{cyellow}e_{a}^{\gamma}}e_{i_{3}}^{\gamma}\varphi_{i_{1}}\varphi_{i_{2}}\varphi_{i_{3}}\notag\\
 & = & \bar{\lambda}_{3}^{3}\left[(n+1)\sum_{\alpha\beta}{\color{cyellow}e_{b}^{\alpha}}e_{i_{1}}^{\alpha}{\color{cyellow}e_{b}^{\beta}}e_{c}^{\beta}e_{i_{2}}^{\beta}e_{c}^{\alpha}e_{i_{3}}^{\alpha}-\sum_{\alpha\beta\gamma}{\color{cyellow}e_{b}^{\alpha}}e_{i_{1}}^{\alpha}{\color{cyellow}e_{b}^{\beta}}e_{c}^{\beta}e_{i_{2}}^{\beta}e_{c}^{\gamma}e_{i_{3}}^{\gamma}\right]\varphi_{i_{1}}\varphi_{i_{2}}\varphi_{i_{3}}\notag\\
 & = & \bar{\lambda}_{3}^{3}\left[(n+1)^{2}\sum_{\alpha}e_{i_{1}}^{\alpha}{\color{cyellow}e_{c}^{\alpha}}e_{i_{2}}^{\alpha}{\color{cyellow}e_{c}^{\alpha}}e_{i_{3}}^{\alpha}-(n+1)\sum_{\alpha\beta}e_{i_{1}}^{\alpha}{\color{cyellow}e_{c}^{\beta}}e_{i_{2}}^{\beta}{\color{cyellow}e_{c}^{\alpha}}e_{i_{3}}^{\alpha}\right.\notag\\
 &  & \left.-(n+1)\sum_{\alpha\gamma}e_{i_{1}}^{\alpha}{\color{cyellow}e_{c}^{\alpha}}e_{i_{2}}^{\alpha}{\color{cyellow}e_{c}^{\gamma}}e_{i_{3}}^{\gamma}+\sum_{\alpha\beta\gamma}{\color{red}e_{i_{1}}^{\alpha}}e_{c}^{\beta}e_{i_{2}}^{\beta}e_{c}^{\gamma}e_{i_{3}}^{\gamma}\right]\varphi_{i_{1}}\varphi_{i_{2}}\varphi_{i_{3}}\notag\\
 & = & \bar{\lambda}_{3}^{3}\left[n(n+1)^{2}\sum_{\alpha}e_{i_{1}}^{\alpha}e_{i_{2}}^{\alpha}e_{i_{3}}^{\alpha}-(n+1)^{2}\sum_{\alpha}e_{i_{1}}^{\alpha}e_{i_{2}}^{\alpha}e_{i_{3}}^{\alpha}+\right.\notag\\
 &  & \left.+(n+1)\sum_{\alpha\beta}e_{i_{1}}^{\alpha}{\color{red}e_{i_{2}}^{\beta}}e_{i_{3}}^{\alpha}-(n+1)^{2}\sum_{\alpha}e_{i_{1}}^{\alpha}e_{i_{2}}^{\alpha}e_{i_{3}}^{\alpha}+(n+1)\sum_{\alpha\beta\gamma}e_{i_{1}}^{\alpha}e_{i_{2}}^{\alpha}{\color{red}e_{i_{3}}^{\gamma}}\right]\varphi_{i_{1}}\varphi_{i_{2}}\varphi_{i_{3}}\notag\\
 & = & \bar{\lambda}_{3}^{3}\left[n(n+1)^{2}\sum_{\alpha}e_{i_{1}}^{\alpha}e_{i_{2}}^{\alpha}e_{i_{3}}^{\alpha}-2(n+1)^{2}\sum_{\alpha}e_{i_{1}}^{\alpha}e_{i_{2}}^{\alpha}e_{i_{3}}^{\alpha}\right]\varphi_{i_{1}}\varphi_{i_{2}}\varphi_{i_{3}}\notag\\
 & = & \bar{\lambda}_{3}^{3}(n+1)^{2}(n-2)I_3\,.
\end{eqnarray}
Inserting these traces into \eqref{eq:flow2} and comparing with the flow equation in the form \eqref{eq:betaflow} we can immediately read off the  (dimensionful) beta functions
$$
\bar{\beta}_{2}=k^{d-4}\frac{2(n-1)(n+1)^{2}}{(1+\lambda_{2})^{3}}\bar{\lambda}_{3}^{2}
\qquad\qquad
\bar{\beta}_{3}=-k^{d-6}\frac{6(n-2)(n+1)^{2}}{(1+\lambda_{2})^{4}}\bar{\lambda}_{3}^{3}\,.
$$

\subsubsection{Explicit beta functions}\label{sec:fullbetan}

It is clear form the previous example that the computation of the general trace ${\rm tr }\,{\mathbb M}^m$ can be performed along the same lines but rapidly becomes unfeasible by hand, and for a given truncation order $p$ in the expansion \eqref{eq:Vexp} we need to expand the inverse propagator up to ${\rm tr}\,\mathbb{M}^{p}$ and keep all contributions up to order $p$. 
To tackle the "trace machinery" involved in the general computation we have  used symbolic manipulation software (the {\tt xTensor} package for {\tt Mathematica} \cite{xTensor}) for which we have written an explicit code.

In this work we consider the expansion \eqref{eq:Vexp} up to order $\varphi^{6}$ for a total of ten dimensionless couplings:
$\left\{\lambda_2,\lambda_3,\lambda_{4,1},\lambda_{4,2},\lambda_{5,1},\lambda_{5,2},\lambda_{6,1},\lambda_{6,2},\lambda_{6,3},\lambda_{6,4}\right\}$.
%In order to implement the similarity of the RG transformation and therefore to ensure scale invariance of the effective action $\Gamma_k$ at the critical point 
The beta functions of these couplings can be written in terms of the dimensionfull ones as\footnote{We just differentiate both sides of $\lambda_{m,i}=k^{m(d/2-1+\eta/2)-d}\bar\lambda_{m,i}$.}
%
%$$
%\beta_m=k^{d-m(d/2-1+\eta_k/2)}\left\{[m(d/2-1+\eta_k/2)-d]\lambda_m+\bar{\beta}_m\right\}\,.
%$$
%
%
\begin{equation}
\beta_{m,i}=\left[m\left(\frac{d}{2}-1+\frac{\eta}{2}\right)-d\right]\lambda_{m,i}+k^{m\left(\frac{d}{2}-1+\frac{\eta}{2}\right)-d}\,\bar\beta_{m,i}\,,
\end{equation}
and their explicit form %The corresponding system of (dimensionless) beta functions 
can be extracted once the reduction of the traces up to ${\rm  tr} \, \mathbb{M}^6$ has been performed.
The final result is the following set of ten beta functions, valid for arbitrary $d$ and $n$, and it is the main achievement of the present work:
{\small
\begin{eqnarray}
\beta_{2}  &=&  (-2+\eta)\lambda_{2}+\frac{2(n-1)(n+1)^{2}\lambda_{3}^{2}}{(1+\lambda_{2})^{3}}
 \nonumber\\&&
-\frac{n(n+1)\lambda_{4,1}+\frac{1}{3}(n+2)\lambda_{4,2}}{(1+\lambda_{2})^{2}}
 \label{b2}
\end{eqnarray}
\begin{eqnarray}
\beta_{3} & = & \frac{1}{2}(d+3\eta-6)\lambda_{3}+\frac{6\!\left(\frac{2}{3}\lambda_{4,2}+(n-1)(n+1)\lambda_{4,1}\right)\!\lambda_{3}}{(1+\lambda_{2})^{3}}\nonumber\\
 &  & -\frac{n\lambda_{5,1}+\frac{1}{10}(n+6)\lambda_{5,2}}{(1+\lambda_{2})^{2}}-\frac{6(n-2)(n+1)^{2}\lambda_{3}^{3}}{(1+\lambda_{2})^{4}}
 \label{b3}
\end{eqnarray}
\begin{eqnarray}
\beta_{4,1} & = & (d+2\eta-4)\lambda_{4,1}+\frac{6\!\left(\frac{4}{3}\lambda_{4,2}\lambda_{4,1}+(n^{2}-1)\lambda_{4,1}^{2}\right)}{(1+\lambda_{2})^{3}}\nonumber\\
 &  & +\frac{6\frac{4}{15}(n+1)\!\left(3\lambda_{5,2}+5(n-1)\lambda_{5,1}\right)\!\lambda_{3}}{(1+\lambda_{2}){}^{3}}\nonumber\\
 &  & -\frac{24\!\left(\frac{3}{2}(n-2)(n+1)^{2}\lambda_{4,1}+(n+1)\lambda_{4,2}\right)\!\lambda_{3}^{2}}{(1+\lambda_{2})^{4}}\nonumber\\
 &  & -\frac{15n\lambda_{6,1}+(n+8)\lambda_{6,3}+9(n+1)\lambda_{6,2}}{15(1+\lambda_{2}){}^{2}}\nonumber\\
 &  & +\frac{24(n-3)(n+1)^{3}\lambda_{3}^{4}}{(1+\lambda_{2}){}^{5}}
 \label{b41}
\end{eqnarray}
\begin{eqnarray}
\beta_{4,2} & = & (d+2\eta-4)\lambda_{4,2}-\frac{24\!\left(\frac{3}{2}(n+1)^{3}\lambda_{4,1}+\frac{1}{2}(n-3)(n+1)^{2}\lambda_{4,2}\right)\!\lambda_{3}^{2}}{(1+\lambda_{2})^{4}}\nonumber\\
 &  & +\frac{6(n+1)^{2}\lambda_{4,1}^{2}+6\frac{2}{3}n(n+1)\lambda_{4,1}\lambda_{4,2}+6\frac{1}{9}(n+8)\lambda_{4,2}^{2}}{(1+\lambda_{2})^{3}}\nonumber\\
 &  & +\frac{\frac{12}{5}(n-3)(n+1)^{2}\lambda_{3}\lambda_{5,2}}{(1+\lambda_{2})^{3}}\nonumber\\
 &  & +\frac{3(n+1)^{2}\lambda_{6,2}-2n(n+1)\lambda_{6,3}-(n+4)\lambda_{6,4}}{5(1+\lambda_{2})^{2}}\nonumber\\
 &  & +\frac{48(n+1)^{4}\lambda_{3}^{4}}{(1+\lambda_{2}){}^{5}}
  \label{b42}
\end{eqnarray}
\begin{eqnarray}
\beta_{5,1} & = & \frac{1}{2}(3d+5\eta-10)\lambda_{5,1}+\frac{80(n+1)^{2}\!\left(3(n+1)(n-3)\lambda_{4,1}+2\lambda_{4,2}\right)\!\lambda_{3}^{3}}{(1+\lambda_{2})^{5}}\nonumber\\
 &  & -\frac{6(n+1)\!\left(15(n+1)(n-2)\lambda_{4,1}^{2}+20\lambda_{4,1}\lambda_{4,2}\right)\!\lambda_{3}}{(1+\lambda_{2})^{4}}\nonumber\\
 &  & -\frac{12(n+1)^{2}\!\left(5(n-2)\lambda_{5,1}+3\lambda_{5,2}\right)\!\lambda_{3}^{2}}{(1+\lambda_{2})^{4}}\nonumber\\
 &  & +\frac{\frac{4}{3}\!\left(10\lambda_{4,2}\lambda_{5,1}+9(n+1)\lambda_{4,1}\lambda_{5,2}+15(n+1)(n-1)\lambda_{4,1}\lambda_{5,1}\right)}{(1+\lambda_{2})^{3}}\nonumber\\
 &  & +\frac{\frac{2}{3}(n+1)\!\left(8\lambda_{6,3}+15(n-1)\lambda_{6,1}+9(n+1)\lambda_{6,2}\right)\!\lambda_{3}}{(1+\lambda_{2})^{3}}\nonumber\\
 &  & -\frac{120(n-4)(n+1)^{4}\lambda_{3}^{5}}{(1+\lambda_{2}){}^{6}}
  \label{b51}
\end{eqnarray}
\begin{eqnarray}
\beta_{5,2} & = & \frac{1}{2}(3d+5\eta-10)\lambda_{5,2}-\frac{20\left(9(n+1)^{2}\lambda_{4,1}^{2}+3\left(n^{2}-2n-3\right)\lambda_{4,2}\lambda_{4,1}+4\lambda_{4,2}^{2}\right)\!\lambda_{3}}{(1+\lambda_{2}){}^{4}}\nonumber\\
 &  & -\frac{6(n+1)^{2}\left(10\lambda_{5,1}+(4n-19)\lambda_{5,2}\right)\!\lambda_{3}^{2}}{(1+\lambda_{2}){}^{4}}\nonumber\\
 &  & +\frac{\frac{4}{3}\left(6\lambda_{6,4}+\left(3n^{2}-4n-7\right)\lambda_{6,3}+3(n-4)(n+1)^{2}\lambda_{6,2}\right)\!\lambda_{3}}{(1+\lambda_{2}){}^{3}}\nonumber\\
 &  & +\frac{2(n+1)\left(10\lambda_{5,1}+(4n-9)\lambda_{5,2}\right)\lambda_{4,1}+\left(26\lambda_{5,2}+n\left(10\lambda_{5,1}+\lambda_{5,2}\right)\right)\lambda_{4,2}}{(1+\lambda_{2}){}^{3}}\nonumber\\
 &  & +\frac{80(n+1)^{2}\left(9(n+1)\lambda_{4,1}+(n-6)\lambda_{4,2}\right)\!\lambda_{3}^{3}}{(1+\lambda_{2}){}^{5}}-\frac{600(n+1)^{4}\lambda_{3}^{5}}{(1+\lambda_{2}){}^{6}}
  \label{b52}
\end{eqnarray}
\begin{eqnarray}
\beta_{6,1} & = & \left(2d+3\eta-6\right)\lambda_{6,1}-\frac{600(n+1)^{3}\left(3\left(n^{2}-3n-4\right)\lambda_{4,1}+2\lambda_{4,2}\right)\!\lambda_{3}^{4}}{(1+\lambda_{2}){}^{6}}\nonumber\\
 &  & +\frac{24(n+1)^{2}\left(60\lambda_{4,1}\lambda_{4,2}+45\left(n^{2}-2n-3\right)\lambda_{4,1}^{2}\right)\!\lambda_{3}^{2}}{(1+\lambda_{2}){}^{5}}\nonumber\\
 &  & +\frac{96(n+1)^{3}\left(5(n-3)\lambda_{5,1}+3\lambda_{5,2}\right)\!\lambda_{3}^{3}}{(1+\lambda_{2}){}^{5}}\nonumber\\
 &  & -\frac{24(n+1)\left(3(n+1)\left(5(n-2)\lambda_{5,1}+3\lambda_{5,2}\right)\lambda_{4,1}+10\lambda_{4,2}\lambda_{5,1}\right)\!\lambda_{3}}{(1+\lambda_{2}){}^{4}}\nonumber\\
 &  & -\frac{90(n+1)\left(\left(n^{2}-n-2\right)\lambda_{4,1}+2\lambda_{4,2}\right)\!\lambda_{4,1}^{2}}{(1+\lambda_{2}){}^{4}}\nonumber\\
 &  & -\frac{6(n+1)^{2}\left(15(n-2)\lambda_{6,1}+8\lambda_{6,3}+9(n+1)\lambda_{6,2}\right)\!\lambda_{3}^{2}}{(1+\lambda_{2}){}^{4}}\nonumber\\
 &  & +\frac{20\lambda_{4,2}\lambda_{6,1}+4(n+1)\lambda_{5,1}\left(5(n-1)\lambda_{5,1}+6\lambda_{5,2}\right)}{(1+\lambda_{2}){}^{3}}\nonumber\\
 &  & +\frac{2(n+1)\left(15(n-1)\lambda_{6,1}+8\lambda_{6,3}+9(n+1)\lambda_{6,2}\right)\!\lambda_{4,1}}{(1+\lambda_{2}){}^{3}}\nonumber\\
 &  & +\frac{720(n-5)(n+1)^{5}\lambda_{3}^{6}}{(1+\lambda_{2}){}^{7}}
  \label{b61}
\end{eqnarray}
\begin{eqnarray}
\beta_{6,2} & = & \left(2d+3\eta-6\right)\lambda_{6,2}+\frac{20\left(5\lambda_{4,2}\lambda_{6,2}+\lambda_{4,1}\left(-4\lambda_{6,3}+3\left(n^{2}-1\right)\lambda_{6,2}\right)\right)}{5(1+\lambda_{2}){}^{3}}\nonumber\\
 &  & +\frac{100\lambda_{5,1}^{2}+20n\lambda_{5,1}\lambda_{5,2}+(n+30)\lambda_{5,2}^{2}}{5(1+\lambda_{2}){}^{3}}\nonumber\\
 &  & -\frac{1200(n+1)^{2}\left(3(n+1)\lambda_{4,1}-\lambda_{4,2}\right)\!\lambda_{3}^{4}}{(1+\lambda_{2}){}^{6}}\nonumber\\
 &  & +\frac{48(n+1)^{2}\left(10\lambda_{5,1}+(n-8)\lambda_{5,2}\right)\lambda_{3}^{3}}{(1+\lambda_{2}){}^{5}}\nonumber\\
 &  & +\frac{16\left(15(n+1)\left(3(n+1)\lambda_{4,1}-4\lambda_{4,2}\right)\lambda_{4,1}+5\left(9(n+1)^{2}\lambda_{4,1}^{2}+2\lambda_{4,2}^{2}\right)\right)\lambda_{3}^{2}}{(1+\lambda_{2}){}^{5}}\nonumber\\
 &  & -\frac{6\left(-30\lambda_{4,1}^{2}\lambda_{4,2}+6(n+1)\left(10\lambda_{5,1}+(n-4)\lambda_{5,2}\right)\!\lambda_{3}\lambda_{4,1}\right)}{(1+\lambda_{2}){}^{4}}\nonumber\\
 &  & -\frac{6\left(3(n+1)^{2}(2n-7)\lambda_{3}\lambda_{6,2}-8\left((n+1)\lambda_{3}\lambda_{6,3}-2\lambda_{4,2}\lambda_{5,2}\right)\right)\lambda_{3}}{(1+\lambda_{2}){}^{4}}\nonumber\\
 &  & +\frac{2160(n+1)^{4}\lambda_{3}^{6}}{(1+\lambda_{2}){}^{7}}
  \label{b62}
\end{eqnarray}
\begin{eqnarray}
\beta_{6,3} & = & \left(2d+3\eta-6\right)\lambda_{6,3}+\frac{16(n+1)\left(60(n+1)^{2}\lambda_{5,1}-63\lambda_{5,2}\right)\!\lambda_{3}^{3}}{(1+\lambda_{2}){}^{5}}\nonumber\\
 &  & +\frac{16(n+1)\left(9n^{3}-45n^{2}-117n\right)\!\lambda_{3}^{3}\lambda_{5,2}}{(1+\lambda_{2}){}^{5}}\nonumber\\
 &  & +\frac{16(n+1)\left(180(n+1)^{2}\lambda_{4,1}^{2}+15\left(3n^{2}-11n-14\right)\lambda_{4,1}\lambda_{4,2}+50\lambda_{4,2}^{2}\right)\!\lambda_{3}^{2}}{(1+\lambda_{2}){}^{5}}\nonumber\\
 &  & -\frac{6(n+1)\left(7\left(n^{2}-2n-3\right)\lambda_{6,3}-18(n+1)^{2}\lambda_{6,2}\right)\!\lambda_{3}^{2}}{(1+\lambda_{2}){}^{4}}\nonumber\\
 &  & -\frac{18(n+1)\left(5(n+1)\lambda_{6,1}+4\lambda_{6,4}\right)\!\lambda_{3}^{2}}{(1+\lambda_{2}){}^{4}}\nonumber\\
 &  & -\frac{30\left(9(n+1)^{2}\lambda_{4,1}^{2}+3\left(n^{2}-1\right)\lambda_{4,2}\lambda_{4,1}+8\lambda_{4,2}^{2}\right)\!\lambda_{4,1}}{(1+\lambda_{2}){}^{4}}\nonumber\\
 &  & -\frac{12(n+1)\left(2\left(5(n-3)\lambda_{5,1}+9\lambda_{5,2}\right)\lambda_{4,2}+3(n+1)\left(10\lambda_{5,1}+3(n-5)\lambda_{5,2}\right)\lambda_{4,1}\right)\!\lambda_{3}}{(1+\lambda_{2}){}^{4}}\nonumber\\
 &  & +\frac{18(n+1)\left(10(n-3)\lambda_{5,1}+9\lambda_{5,2}\right)\lambda_{5,2}}{15(1+\lambda_{2}){}^{3}}\nonumber\\
 &  & +\frac{10\left(15n\lambda_{6,1}+(n+38)\lambda_{6,3}+9(n+1)\lambda_{6,2}\right)\!\lambda_{4,2}}{15(1+\lambda_{2}){}^{3}}\nonumber\\
 &  & +\frac{30\left(15(n+1)\lambda_{6,1}+\left(7n^{2}+n-6\right)\lambda_{6,3}+12\lambda_{6,4}-18(n+1)^{2}\lambda_{6,2}\right)\!\lambda_{4,1}}{15(1+\lambda_{2}){}^{3}}\nonumber\\
 &  & -\frac{600(n+1)^{3}\left(12(n+1)\lambda_{4,1}+(n-7)\lambda_{4,2}\right)\!\lambda_{3}^{4}}{(1+\lambda_{2}){}^{6}}\nonumber\\
 &  & +\frac{4320(n+1)^{5}\lambda_{3}^{6}}{(1+\lambda_{2}){}^{7}}
  \label{b63}
\end{eqnarray}
\begin{eqnarray}
\beta_{6,4} & = & \left(2d+3\eta-6\right)\lambda_{6,4}+\frac{600(n+1)^{4}\left(3(n+1)\lambda_{4,1}-4\lambda_{4,2}\right)\!\lambda_{3}^{4}}{(1+\lambda_{2}){}^{6}}\nonumber\\
 &  & +\frac{8(n+1)^{2}\left(-90(n+1)^{2}\lambda_{4,1}^{2}+72(n+1)^{2}\lambda_{3}\lambda_{5,2}\right)\!\lambda_{3}^{2}}{(1+\lambda_{2}){}^{5}}\nonumber\\
 &  & +\frac{8(n+1)^{2}\left(150(n+1)\lambda_{4,1}\lambda_{4,2}+5(3n-23)\lambda_{4,2}^{2}\right)\!\lambda_{3}^{2}}{(1+\lambda_{2}){}^{5}}\nonumber\\
 &  & -\frac{108(n+1)^{2}\left(6(n+1)\lambda_{4,1}+(n-7)\lambda_{4,2}\right)\,\lambda_{3}\lambda_{5,2}}{3(1+\lambda_{2}){}^{4}}\nonumber\\
 &  & -\frac{54(n+1)^{2}\left(2(n+1)\lambda_{6,3}+(n-5)\lambda_{6,4}+3(n+1)^{2}\lambda_{6,2}\right)\!\lambda_{3}^{2}}{3(1+\lambda_{2}){}^{4}}\nonumber\\
 &  & -\frac{10\left(-27(n+1)^{3}\lambda_{4,1}^{3}+27(n+1)^{2}\lambda_{4,1}^{2}\lambda_{4,2}+9n(n+1)\lambda_{4,1}\lambda_{4,2}^{2}+(n+26)\lambda_{4,2}^{3}\right)}{3(1+\lambda_{2}){}^{4}}\nonumber\\
 &  & +\frac{2\left(6(n+1)^{2}\lambda_{4,1}\lambda_{6,3}+9(n+1)^{3}\lambda_{4,1}\lambda_{6,2}+14\lambda_{4,2}\lambda_{6,4}\right)}{(1+\lambda_{2}){}^{3}}\nonumber\\
 &  & +\frac{2\left(3n(n+1)\lambda_{4,1}\lambda_{6,4}+2n(n+1)\lambda_{4,2}\lambda_{6,3}+n\lambda_{4,2}\lambda_{6,4}-3(n+1)^{2}\lambda_{4,2}\lambda_{6,2}\right)}{(1+\lambda_{2}){}^{3}}\nonumber\\
 &  & +\frac{9(n-7)(n+1)^{2}\lambda_{5,2}^{2}}{5(1+\lambda_{2}){}^{3}}-\frac{1440(n+1)^{6}\lambda_{3}^{6}}{(1+\lambda_{2}){}^{7}}\,.
 \label{b64}
\end{eqnarray}
}
\\
\!\!The anomalous dimension $\eta$ entering these expressions will be computed in the next subsection.

While the beta functions \eqref{b2}--\eqref{b64} are written using the linear cutoff, it is easy
to shift to a general cutoff $R_k(z)$ by the substitution
$$
\frac{1}{(1+\lambda_{2})^{m}}\,\to\,k^{2m-2-d} \,\frac{d}{4} \int_0^{\infty}dz\,z^{d/2-1}G_k^m(z)\partial_tR_k(z)\,.
$$
In this way it is straightforward to obtain beta functions for an arbitrary cutoff as needed for the study of cutoff dependence and for the optimization of  convergence. We leave this task to a future study, since, as we will show in section \ref{GA}, within the $p=6$ truncation convergence of critical exponents is fully achieved only in $d=5$ while larger truncations are needed in $d=4$ and in $d=3$.

\subsection{Anomalous dimension}
The computation of the anomalous dimension $\eta_k$ requires the computation of the flow of the wavefunction $Z_k$ since $\eta_k=-\partial_t\log Z_k$.
%As for the effective potential, this is possible only after we define $Z_k$ in terms of the effective action $\Gamma_k$ itself. 
It is clear from \eqref{eq:ansatz} that $Z_k$ corresponds to the term in $\Gamma_k$ which is quadratic in the fields and in the momentum
\begin{equation}\label{eq:defZ}
\partial_t Z_k \,\delta_{ij}=\left.\lim_{p^2\to 0}\frac{d^2}{dp^2}\frac{\delta^{2}}{\delta\varphi_i(p)\delta\varphi_j(-p)} \partial_t\Gamma_k[\varphi]\right|_{\varphi=0}\,.
\end{equation} 
The flow of $Z_k$ is therefore related to that of the two-point function whose flow equation reads % \cite{} (aggiungere $ij$)
\begin{equation}
[\partial_{t}\Gamma_{k}^{(2)}(p^{2})]_{ij}  = -\frac{1}{2}\int_{q}[\Gamma_{k}^{(4)}(q,p,-p,-q)]_{aija}G_{k}^2(q^{2})\partial_{t}R_{k}(q^{2}) \nonumber
\end{equation}
\begin{equation}\label{eq:2pointflow} +\int_{q}[\Gamma_{k}^{(3)}(q,p,-q-p)]_{aib} G_{k}((q+p)^{2}) [\Gamma_{k}^{(3)}(q+p,-p,-q)]_{bja}G_{k}^2(q^{2})\partial_{t}R_{k}(q^{2})\,,
\end{equation}
where we introduced the regularized propagator (at $\varphi=0$)
$$
[G_k(q^2)]^{-1}= \Gamma_k^{(2)}(q^2)+R_k(q^2) = Z_k q^2+\bar{\lambda}_2+R_k(q^2)\,.
$$
Equation \eqref{eq:2pointflow} depends on the three- and four-point functions, but the only contribution proportional to $p^2$ comes from the integral involving the first. Thus without loss of generality we can consider the effective action \eqref{eq:ansatz} where the potential \eqref{eq:Vexp} is truncated at order $p=3$ finding
\begin{equation}
\partial_t Z_k \delta_{ij} = \bar{\lambda}_{3}^{2} \, T_{aib}^{(3)}T_{ajb}^{(3)} \left.\int_{q}G_{k}((q+p)^{2})G_{k}^2(q^{2})\partial_{t}R_{k}(q^{2})\right|_{p^2} \,.% = \frac{k^{d+2}}{(Z_k k^2+\bar{\lambda}_2)^4}\,.
\label{dtZ}
\end{equation}
Employing the linear cutoff (including now the wavefunction) $R_k(z) = Z_k (k^2-z)\theta(k^2-z)$ gives
%The terms proportional to $p^2$ on the  the r.h.s. of eq. \eqref{eq:2pointflow}, we proceed expanding the regularized propagator
%
%\begin{equation}
%\begin{split}
%G_k((q+p)^2)&\equiv G_k(q^2+p^2+2qpx)\\
%&=G_k(q^2)+(p^2+2qpx)G_k'(q^2)+\\
%&+\frac{1}{2}(p^4+4p^2q^2x^2+4p^3qx)G_k''(q^2)+O(p^3)
%\end{split}
%\end{equation}
%%
%and though this general analysis is valid for any cutoff function, in the following we choose again the linear cutoff introduced in the previous section. With the help of the following relations
%With the linear cutoff we find:
%%
%\begin{eqnarray*}
%R'_k(q^2) & = & -Z_k\theta(k^2-q^2)\\
%R''_k(q^2) & = & Z_k\delta(k^2-q^2)\\
%\partial_tR_k(q^2) & = & Z_k[(2-\eta)k^2+\eta q^2]\theta(k^2-q^2)\\
%G_k'(q^2) & = & -(Z_k+R'_k(q^2))G_k(q^2)^2\\
%G_k''(q^2) & = & -(Z_k+R'_k(q^2))^2G_k(q^2)^3-G_k(q^2)^2R_k''(q^2)
%\end{eqnarray*}
%%
%the corresponding integral on the r.h.s. of \eqref{eq:2pointflow} reads 
%
\begin{equation}
\left.\int_{q}G_{k}((q+p)^{2})G_{k}^2(q^{2})\partial_{t}R_{k}(q^{2})\right|_{p^2}  = -Z_k^2c_d \left(1-\frac{\eta_k}{d+2} \right)\frac{k^{d+2}}{(Z_k k^2+\bar{\lambda}_2)^4}\,.
\end{equation}
At this point we are left with the task of computing the trace $T_{aib}^{(3)}T_{ajb}^{(3)}$ which, on the other hand, we already computed in section \ref{ssec:phi3} when calculating $\text{tr}\,\mathbb{M}^2$ in \eqref{eq:trm2}. Equation \eqref{dtZ} then becomes
\begin{equation*}
\partial_{t}Z_k  
%=  \frac{\bar{\lambda}{}_{3}^{2} k^{d+2}}{(Z_k k^2+\bar{\lambda}_2)^4}T_{aib}^{(3)}T_{ajb}^{(3)}
= -Z_k^2c_d\left(1-\frac{\eta_k}{d+2} \right)\frac{k^{d+2}}{(Z_k k^2+\bar{\lambda}_2)^4} (n-1)(n+1)^{2}\, \bar{\lambda}_{3}^{2} \,.
\end{equation*}
Finally, switching to dimensionless variables $\bar{\lambda}_m = Z_k^{m/2}k^{d-m (d/2-1)}\lambda_m$ and after re-absorbing the factor $c_d$ in a field redefinition as before, we find the following simple expression for the anomalous dimension
\begin{equation}\label{eq:andim}
\eta_k=(n-1)(n+1)^{2}\frac{\lambda_{3}^{2}}{(1+\lambda_{2})^{4}} \left(1-\frac{\eta_k}{d+2} \right)\,.
\end{equation}
Even if the anomalous dimension term on the r.h.s of this equation steams from the non-perturbative part of the flow equation, we will omit it in the follwoing since its contributions turns out to be negligible in $d=4,5$ since $\eta$ is small in that range of dimensions.  
We remark here that in the LPA' the anomalous dimension receives contributions only from the three-point function at $\varphi=0$, i.e. from  the trilinear coupling $\lambda_3$, irrespectively of the number of LPA' couplings considered: we therefore expect no further corrections to \eqref{eq:andim} when the potential \eqref{eq:Vexp} is truncated at higher orders. Beyond the LPA' approximation scheme instead, the expression for the anomalous dimension would receive additional contributions, but this will not be considered in this work.
\subsection{LPA' at fixed $n$}\label{LPA}
The flow equation \eqref{eq:flow} can be expressed exactly, in the lowest cases,
using the Cayley\textendash Hamilton theorem to compute $(k^{2} \mathbb{I}+\mathbb{V})^{-1}$. For $n=1,2,3$ we find the following explicit forms
\begin{align}
\partial_{t}V_{k}&=k^{d+2}\frac{1}{k^{2}+{\rm tr}\mathbb{V}}&&n=1\label{eq:LPA1}\\
\partial_{t}V_{k}&=k^{d+2}\frac{2k^{2}+{\rm tr}\mathbb{V}}{k^{4}+k^{2}{\rm tr}\mathbb{V}+\det\mathbb{V}}&&n=2\label{eq:LPA2}\\
\partial_{t}V_{k}&=k^{d+2}\frac{3k^{4}+2k^{2}{\rm tr}\mathbb{V}-\frac{1}{2}\left({\rm tr}(\mathbb{V}^{2})-({\rm tr}\mathbb{V})^{2}\right)}{k^{6}+k^{4}{\rm tr}\mathbb{V}-\frac{1}{2}k^{2}\left({\rm tr}(\mathbb{V}^{2})-({\rm tr}\mathbb{V})^{2}\right)+\det\mathbb{V}}&&n=3\label{eq:LPA3}\,.
\end{align}
%
%\begin{equation}\label{eq:LPA1}
%\partial_{t}V_{k}\overset{n=1}{=}k^{d+2}\frac{1}{k^{2}+{\rm tr}\mathbb{V}}\,,
%\end{equation}
%% 
%\begin{equation}\label{eq:LPA2}
%\partial_{t}V_{k}\overset{n=2}{=}k^{d+2}\frac{2k^{2}+{\rm tr}\mathbb{V}}{k^{4}+k^{2}{\rm tr}\mathbb{V}+\det\mathbb{V}}
%\end{equation}
%%
%and
%%
%\begin{equation}\label{eq:LPA3}
%\partial_{t}V_{k}\overset{n=3}{=}k^{d+2}\frac{3k^{4}+2k^{2}{\rm tr}\mathbb{V}-\frac{1}{2}\left({\rm tr}(\mathbb{V}^{2})-({\rm tr}\mathbb{V})^{2}\right)}{k^{6}+k^{4}{\rm tr}\mathbb{V}-\frac{1}{2}k^{2}\left({\rm tr}(\mathbb{V}^{2})-({\rm tr}\mathbb{V})^{2}\right)+\det\mathbb{V}}\,.
%\end{equation}
%
Up to now these are just the flow equations for an $n$-component scalar
$\varphi_{i}$ since the symmetry has not been imposed yet.
The information relative to the discrete symmetry $S_{n+1}$ enters through the form of the invariants which the potential is function of.
In general, the LPA' for $S_{n+1}$ depends on $n$ independent invariants, as can be seen explicitly for the $n=1,2,3$ cases in Table \ref{T_inv2}. In what follows any $k$ dependence is intended.

In the {\tt{Ising}} case $n=1$ the only invariant is $\rho=\varphi_{1}^{2}$
so we define $U(\rho)\equiv V(\varphi_{1})$.  Since by change of variables, the second functional derivative of the effective potential w.r.t. $\varphi_1$ reads $V_{11}=2U_{\rho}+4\rho U_{\rho\rho}$,
the LPA' \eqref{eq:LPA1} takes the well known form
\begin{equation}\label{eq:IsingLPA}
\partial_{t}U=\frac{k^{d+2}}{k^{2}+2U_{\rho}+4\rho U_{\rho\rho}}\,.
\end{equation}
Standard \texttt{Ising} beta functions are retrieved once we consider,
for example, the following $\varphi^{6}$ truncation
$$
U(\rho)=\frac{1}{2}\bar g_{2}\rho+\frac{1}{4!}\bar g_{4}\rho^{2}+\frac{1}{6!}\bar g_{6}\rho^{3}\,.
$$
Inserting this expression into equation \eqref{eq:IsingLPA}, switching to dimensionless variables and comparing equal powers of the fields gives
\begin{align}
\beta_{2} & =(-2+\eta)g_{2}-\frac{g_{4}}{(1+g_{2})^{2}}\nonumber\\
\beta_{4} & =(d+2\eta-4)g_{4}+\frac{6g_{4}^{2}}{(1+g_{2}){}^{3}}-\frac{g_{6}}{(1+g_{2}){}^{2}}\nonumber\\
\beta_{6} & =(2d+3\eta-6)g_{6}-\frac{90g_{4}^{3}}{(1+g_{2}){}^{4}}+\frac{30g_{6}g_{4}}{(1+g_{2}){}^{3}}\,.
\label{betan1}
\end{align}
We can compare these beta functions against the general beta functions of section
\ref{sec:fullbetan} when $n=1$. Having in mind the result of Table \ref{T_inv2} we find the following mapping between the two representations
$$
g_{2}=\lambda_{2}\qquad\qquad g_{4}=2\lambda_{4,1}+\lambda_{4,2}\qquad\qquad g_{6}=2\lambda_{6,1}+2\lambda_{6,3}+\lambda_{6,4}\,.
$$
If we now make the linear combinations of the general beta functions of section \ref{sec:fullbetan} according to this mapping and set $n=1$, we will find that the r.h.s. of these linear combinations correctly become functions of only the couplings $g_1,g_2,g_3$ and exactly match \eqref{betan1}. This is a non-trivial check of the correctness of our formalism. 
%the general beta functions \eqref{b2},\eqref{b41},\eqref{b42},\eqref{b61},\eqref{b63},\eqref{b64}.

In the same fashion we turn to the  $n=2$ case where, accordingly, we have two invariants 
$$
\rho=\varphi_{1}^{2}+\varphi_{2}^{2}\qquad\qquad\tau=\frac{3}{\sqrt{2}}\varphi_{2}(\varphi_{2}^{2}-3\varphi_{1}^{2})
$$
and the potential is a function of them $U(\rho,\tau)\equiv V(\varphi_{1},\varphi_{2})$.
We can therefore express the flow equation for a two component scalar
field in terms of the $S_{3}$ invariants $(\rho,\tau)$. The derivatives needed to evaluate \eqref{eq:LPA2}
are
\begin{eqnarray*}
V_{11} & = & 2U_{\rho}+4U_{\rho\rho}\varphi_{1}^{2}-9\varphi_{2}\left(\sqrt{2}U_{\tau}+2\varphi_{1}^{2}\left(2\sqrt{2}U_{\rho\tau}-9U_{\tau\tau}\varphi_{2}\right)\right)\\
V_{22} & = & 2U_{\rho}+\frac{81}{2}\left(\varphi_{1}^{2}-\varphi_{2}^{2}\right){}^{2}U_{\tau\tau}+\varphi_{2}\left(4\varphi_{2}U_{\rho\rho}-18\sqrt{2}\left(\varphi_{1}^{2}-\varphi_{2}^{2}\right)U_{\rho\tau}+9\sqrt{2}U_{\tau}\right)
\end{eqnarray*}
and
$$
V_{12}=V_{21}=-\varphi_{1}\varphi_{2}\left(81\left(\varphi_{2}^{2}-\varphi_{1}^{2}\right)U_{\tau\tau}-4U_{\rho\rho}\right)-9\sqrt{2}\varphi_{1}\left(\varphi_{1}^{2}+\varphi_{2}^{2}\right)U_{\rho\tau}+9\sqrt{2}U_{\tau}\,.
$$
Note that these relations are not  yet expressed solely in terms of the invariants. Only the terms $\text{tr}\,\mathbb{V}$ and $\text{det}\,\mathbb{V}$ turn out to be entirely expressible in terms of   $\rho$ and $\tau$
\begin{equation*}
{\rm tr}\,\mathbb{V}=V_{11}+V_{22}=\frac{81\rho^{2}U_{\tau\tau}}{2}+4\rho U_{\rho\rho}+4U_{\rho}+12\tau U_{\rho\tau}\,,
\end{equation*}
\begin{eqnarray*}
\det \mathbb{V} &=& V_{11}V_{22}-(V_{12})^{2} 
\\& =&4U_{\rho}^{2}+U_{\rho}\left(81\rho^{2}U_{\tau\tau}+8\rho U_{\rho\rho}+24\tau U_{\rho\tau}\right)
%\\& 
-6U_{\rho\rho}\left(3\left(2\tau^{2}-9\rho^{3}\right)U_{\tau\tau}+4\tau U_{\tau}\right)\\
 && -18\left(9\rho^{3}-2\tau^{2}\right)U_{\rho\tau}^{2}-324\rho^{2}U_{\rho\tau}U_{\tau}
% \\& 
 -81\rho U_{\tau}\left(3\tau U_{\tau\tau}+2U_{\tau}\right)\,.
\end{eqnarray*}
We can now insert these expressions into the flow equation \eqref{eq:LPA2} for $V(\varphi_{1},\varphi_{2})$
%
%\begin{eqnarray*}
%\partial_{t}V_{k}=\frac{2k^{2}+V_{11}+V_{22}}{\left(k^{2}+V_{11}\right)\left(k^{2}+V_{22}\right)-V_{12}^{2}} & = & \frac{2k^{2}+V_{11}+V_{22}}{k^{4}+k^{2}\left(V_{11}+V_{22}\right)+V_{11}V_{22}-V_{12}^{2}}\,,
%\end{eqnarray*}
to get the flow equation for $U(\rho,\tau)$
%
%{\small
\begin{eqnarray}\label{LPAn2}
\partial_{t}U_{k}\! & = &\!\left\{ \!2k^{2}\!+\!4U_{\rho}\!+\!4\rho U_{\rho\rho}\!+\!12\tau U_{\rho\tau}\!+\!\frac{81}{2}\rho^{2}U_{\tau\tau}\!\right\} \!\!\left\{ k^{4}\!+\!k^{2}\!\left(\!4U_{\rho}\!+\!4\rho U_{\rho\rho}\!+\!12\tau U_{\rho\tau}\!+\!\frac{81}{2}\rho^{2}U_{\tau\tau}\!\right)\right.\nonumber\\
 &  & \!+4U_{\rho}^{2}\!+\!U_{\rho}\!\left(\!8\rho U_{\rho\rho}\!+\!24\tau U_{\rho\tau}\!+\!81\rho^{2}U_{\tau\tau}\!\right)\!-\!18\!\left(\!9\rho^{3}\!-\!2\tau^{2}\!\right)\!\!\left(\!U_{\rho\tau}^{2}\!-\!U_{\tau\tau}U_{\rho\rho}\!\right)\nonumber\\
 &  & \left.\!-\!162\rho U_{\tau}^{2}\!-\!3U_{\tau}\!\left(\!8\tau U_{\rho\rho}\!+\!108\rho^{2}U_{\rho\tau}\!+\!81\rho\tau U_{\tau\tau}\!\right)\!\right\} ^{-1}\,.
\end{eqnarray}
%}
%
This is the explicit form of the $n=2$ LPA'.
As an application we can take advantage of the flow equation in this
form to extract the beta functions of the corresponding $S_{3}$-symmetric
potential. Considering a $\varphi^{6}$ truncation
$$
U(\rho,\tau)=\frac{1}{2!}\bar{g}_{2}\rho+\frac{1}{3!}\bar{g}_{3}\tau+\frac{1}{4!}\bar{g}_{4}\rho^{2}+\frac{1}{5!}\bar{g}_{5}\rho\tau+\frac{1}{6!}(\bar{g}_{6,1}\rho^{3}+\bar{g}_{6,2}\tau^{2})\,,
$$
leads to the following dimensionless beta functions
\begin{eqnarray}
\beta_{2} & = & (-2+\eta)g_{2}+\frac{18g_{3}^{2}}{(1+g_{2}){}^{3}}-\frac{4g_{4}}{3(1+g_{2}){}^{2}}\nonumber\\
\beta_{3} & = & \frac{1}{2}(d+3\eta-6)g_3+\frac{4g_{3}g_{4}}{(1+g_{2}){}^{3}}-\frac{4g_{5}}{5(1+g_{2}){}^{2}}\nonumber\\
\beta_{4} & = & (d+2\eta-4)g_{4}+\frac{972g_{3}^{4}}{(1+g_{2}){}^{5}}-\frac{216g_{3}^{2}g_{4}}{(1+g_{2}){}^{4}}+\frac{648g_{3}g_{5}+100g_{4}^{2}}{15(1+g_{2}){}^{3}}\nonumber\\
 &  & -\frac{12g_{6,1}+27g_{6,2}}{10(1+g_{2}){}^{2}}\nonumber\\
\beta_{5} & = & \frac{1}{2}(3d+5\eta-10)g_{5}+\frac{720g_{3}^{3}g_{4}}{(1+g_{2}){}^{5}}-\frac{216g_{3}^{2}g_{5}}{(1+g_{2}){}^{4}}\nonumber\\
 &  & +\frac{56g_{4}g_{5}-240g_{3}g_{4}^{2}+24g_{3}g_{6,1}+189g_{3}g_{6,2}}{3(1+g_{2}){}^{3}}\nonumber\\
\beta_{6,1} & = & (2d+3\eta-6)g_{6,1}+\frac{131220g_{3}^{6}}{(1+g_{2}){}^{7}}-\frac{48600g_{3}^{4}g_{4}}{(1+g_{2}){}^{6}}+\frac{4680g_{3}^{2}g_{4}^{2}+11664g_{3}^{3}g_{5}}{(1+g_{2}){}^{5}}\nonumber\\
 &  & -\frac{729g_{3}^{2}(4g_{6,1}+9g_{6,2})-560g_{4}^{3}-7776g_{3}g_{4}g_{5}}{6(1+g_{2}){}^{4}}\nonumber\\
 &  & +\frac{5g_{4}(32g_{6,1}+27g_{6,2})+324g_{5}^{2}}{5(1+g_{2}){}^{3}}\nonumber\\
\beta_{6,2} & = & (2d+3\eta-6)g_{6,2}+\frac{160g_{3}^{2}g_{4}^{2}}{(1+g_{2}){}^{5}}-\frac{96g_{3}g_{5}g_{4}}{(1+g_{2}){}^{4}}+\frac{100g_{4}g_{6,2}+32g_{5}^{2}}{5(1+g_{2}){}^{3}}\,.
\label{betan2}
\end{eqnarray}
%
%
%\begin{eqnarray}
%\beta_{2} & = & (-2+\eta)g_{2}+\frac{18g_{3}^{2}}{(1+g_{2}){}^{3}}-\frac{4g_{4}}{3(1+g_{2}){}^{2}}\nonumber\\
%\beta_{3} & = & \frac{1}{2}(d+3\eta-6)+\frac{4g_{3}g_{4}}{(1+g_{2}){}^{3}}-\frac{4g_{5}}{5(1+g_{2}){}^{2}}\nonumber\\
%\beta_{4} & = & (d+2\eta-4)g_{4}+\frac{972g_{3}^{4}}{(1+g_{2}){}^{5}}-\frac{216g_{3}^{2}g_{4}}{(1+g_{2}){}^{4}}+\frac{648g_{3}g_{5}+100g_{4}^{2}}{15(1+g_{2}){}^{3}}
%\nonumber\\ &  &
% -\frac{12g_{6,1}+27g_{6,2}}{10(1+g_{2}){}^{2}}\nonumber
% \end{eqnarray}
% \begin{eqnarray}
%\beta_{5} & = & \frac{1}{2}(3d+5\eta-10)g_{5}+\frac{720g_{3}^{3}g_{4}}{(1+g_{2}){}^{5}}-\frac{216g_{3}^{2}g_{5}}{(1+g_{2}){}^{4}}\nonumber\\
% &  & +\frac{56g_{4}g_{5}-240g_{3}g_{4}^{2}+24g_{3}g_{6,1}+189g_{3}g_{6,2}}{3(1+g_{2}){}^{3}}\nonumber\\
%\beta_{6,1} & = & (2d+3\eta-6)g_{6,1}+\frac{131220g_{3}^{6}}{(1+g_{2}){}^{7}}-\frac{48600g_{3}^{4}g_{4}}{(1+g_{2}){}^{6}}+\frac{4680g_{3}^{2}g_{4}^{2}+11664g_{3}^{3}g_{5}}{(1+g_{2}){}^{5}}\nonumber\\
% &  & -\frac{729g_{3}^{2}(4g_{6,1}+9g_{6,2})-560g_{4}^{3}-7776g_{3}g_{4}g_{5}}{6(1+g_{2}){}^{4}}\nonumber\\
% &  & +\frac{5g_{4}(32g_{6,1}+27g_{6,2})+324g_{5}^{2}}{5(1+g_{2}){}^{3}}\nonumber\\
%\beta_{6,2} & = & (2d+3\eta-6)g_{6,2}+\frac{160g_{3}^{2}g_{4}^{2}}{(1+g_{2}){}^{5}}-\frac{96g_{3}g_{5}g_{4}}{(1+g_{2}){}^{4}}+\frac{100g_{4}g_{6,2}+32g_{5}^{2}}{5(1+g_{2}){}^{3}}\,.
%\label{betan2}
%\end{eqnarray}
%
We can compare this LPA' result against the general beta functions
of section \ref{sec:fullbetan}, now evaluated for $n=2$. Having in mind the results of Table \ref{T_inv2} in the case $n=2$, we get the following mapping between couplings
$$
g_{2}=\lambda_{2}\qquad\qquad g_{3}=\lambda_{3}\qquad\qquad g_{4}=\frac{9}{2}\lambda_{4,1}+\lambda_{4,2}
$$
$$
g_{5}=\frac{5}{2}\lambda_{5,1}+\lambda_{5,2}\qquad\qquad g_{6,1}=\frac{27}{4}\lambda_{6,1}+\frac{9}{2}\lambda_{6,3}+\lambda_{6,4}\qquad\qquad g_{6,2}=\frac{1}{3}\lambda_{6,1}+\lambda_{6,2}\,.
$$
Again, matching the beta functions of section \ref{sec:fullbetan} for $n=2$ with those of \eqref{betan2} according to this mapping works out perfectly, giving us a further strong confirmation of our results.

Finally, it is clear that for any integer $n$ it is possible to find corresponding expressions for the relative LPA', but these flow equations become extremely complicated as $n$ increases.

\section{Applications\label{Applications}}

%We now make a first application of the FRG developed so far.

\subsection{Cubic interaction}

As a first application of the FRG formalism to $S_{n+1}$-symmetric theories,
we consider the simplest case where only a mass term and the trilinear coupling are present.
Within this two couplings truncation we can perform a complete analytical analysis for arbitrary $n$ and $d$.
%Theories with a cubic interaction can be used to model many phase transitions
%{[}...{]} and a naive dimensional analysis for these theories shows
%that around the upper critical dimension $d_{c}=6$ the trilinear
%term dominates over higher orders.

Referring to the general system of dimensionless beta functions of section \ref{sec:fullbetan} we switch off all couplings but $\lambda_2$ and  $\lambda_3$ so that the corresponding dimensionless beta functions \eqref{b2} and \eqref{b3} become
\begin{align*}
\beta_2 & = -2\lambda_2+\eta\lambda_2+2\frac{(n-1)(n+1)^2}{(1+\lambda_2)^3}\lambda_3^2\\
\beta_3 & = \frac{1}{2}\lambda_3(d-6+3\eta)-6\frac{(n-2)(n+1)^2}{(1+\lambda_2)^4}\lambda_3^3\,.
\end{align*}
%
%We notice here that from the functional form of $\beta_{3}$ we expect meaningful solutions
%in the critical range $d<6$ for all those theories with $n\leq2$:
%these are the only ones for which the potential is well defined for
%any $\varphi\in\mathbb{R}$. \textcolor{red}{(more?)}
%
Fixed point solutions $\beta_{i}=0$ give a snapshot of the critical
behavior of such theories in the corresponding theory space. Apart the trivial Gaussian fixed point $(\lambda_{2}^{*}=0,\,\lambda_{3}^{*}=0)$ we find the following non-trivial fixed point
\begin{equation}
\lambda_{2}^{*}  =-\frac{(d-6)(n-1)}{d(n-1)-18n+30}\qquad\qquad
\lambda_{3}^{*}  =\pm\frac{24\sqrt{3}\sqrt{(d-6)(n-2)}(n-2)}{\sqrt{(n+1)^{2}(d(n-1)-18n+30)^{4}}}\,,
\label{fp}
\end{equation}
which, as expected, shrinks to the Gaussian one in $d=6$. 
Note that in $d<6$ and $n>2$ the fixed point is imaginary and as a consequence the theory is non-unitary, as in the case of the {\tt{Lee-Yang}}   universality class which also has $d_c=6$ \cite{zambellizanusso, Codello0, Codello1}. But for $n<2$ the fixed point is real and so is the corresponding Landau-Ginzburg action.  %non-unitary  where there is no such a role played by $n$ to distinguish from a unitary to a non-unitary theory. 

\begin{figure}
\includegraphics[width=1\textwidth]{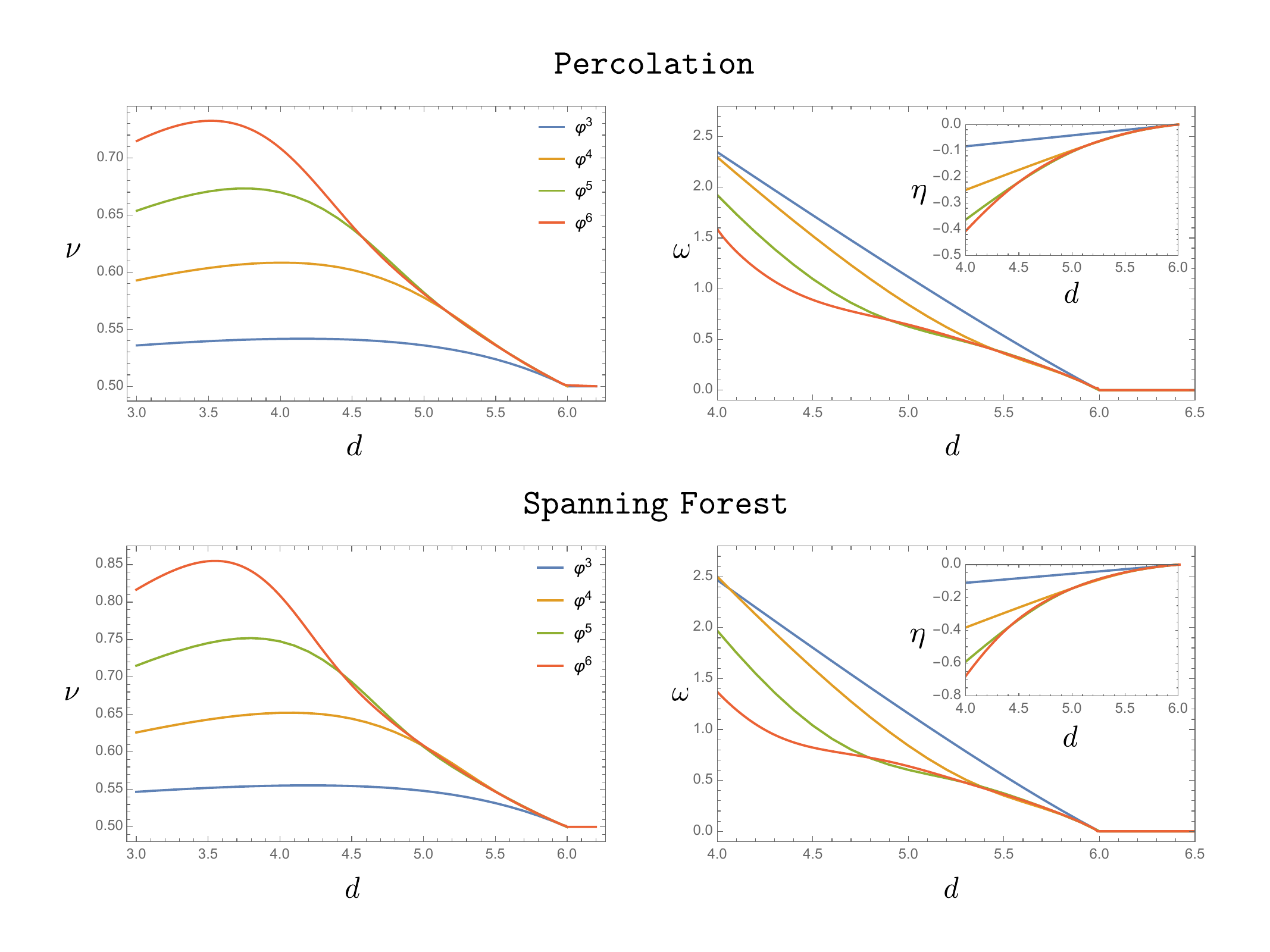}
\caption{{\tt Percolation} (upper plots) and {\tt Spanning Forest} (lower plots) critical exponents $\nu$ (left), $\omega$ (right) and $\eta$ (inset) as a function of $d$ for increasing order of the truncation. For both universality classes, convergence is evident down to $d=5$ while in $d=4$ it is close. In lower dimensions, as in $d=3$, convergence is expected only at higher orders in the truncation.
\label{Fig:nuP}}
\end{figure}

We can now linearize the RG flow around this non-trivial fixed point
to acquire a qualitative understanding of the flow and to extract the critical
exponents. The stability matrix $M_{ij}=\frac{\partial \beta_i}{\partial \lambda_{j}}\Big|_{*}$ around the fixed point has the following components,
\begin{align*}
M_{11} & =-\frac{(d(n-1)-8)(12(n-3)+d(n-1))}{(3n-7)(d-(d-24)n-48)}\\
M_{12} & =\frac{2(6-d)^{3/2}(d-(d-24)n-48)}{3(d(n-1)-12n+20)\sqrt{21-9n}(n+1)}\\
M_{21} & =\frac{24\sqrt{3}(d(n-1)-12n+20)\sqrt{(d-6)(3n-7)}\left(n^{2}-1\right)}{(d-(d-24)n-48)^{2}}\\
M_{22} & =6-d\,,
\end{align*}
with corresponding eigenvalues %$\theta_{\pm}(d,n)$ 
$$
\theta_{\pm}=\frac{1}{2}\left(M_{11}+M_{22}-\sqrt{M_{11}^{2}-2M_{22}M_{11}+M_{22}^{2}+4M_{12}M_{21}}\right)\,.
$$
At this point one can obtain analytical expressions for the critical exponents simply considering that the correlation length critical exponent $\nu$ is related to the inverse of the negative eigenvalue $\nu =-\theta_{-}^{-1}$ while the correction-to-scaling critical exponent $\omega$ can be obtained as the first positive eigenvalue of the stability matrix which in this case is $\omega  =\theta_{+}$.
The anomalous dimension instead is obtained once the fixed point values  \eqref{fp} are substituted into equation \eqref{eq:andim}, obtaining
\begin{equation}
\eta  =\frac{(d-6)(n-1)}{3(7-3n)}\,.
\end{equation}
The critical exponents obtained in this approximation furnish the first curves reported in Figure~\ref{Fig:nuP} in the next subsection. 
%In this case where only the trilinear coupling is considered, the analysis of the critical behavior can be fully carried on analytically. By the way, as soon as more couplings are considered in the expansion \eqref{eq:Vexp}, the beta functions become more complicated and a numerical analysis is called for.

We conclude this section specialising the analysis to $d=6-\epsilon$ dimensions to make contact with the $\epsilon$-expansion. Expanding to first order in $\epsilon$ our results we find 
\begin{equation}
\nu=\frac{1}{2}+\frac{5(n-1)}{12(3n-7)}\epsilon+O(\epsilon^{2})\qquad\qquad\eta=\frac{n-1}{3(7-3n)}\epsilon+O(\epsilon^{2})
\end{equation}
and $\omega = \epsilon+O(\epsilon^{2})$ independently of $n$ \cite{Parisi1998}. These relations are in agreement with \cite{DeAlcantara1980} after a brief manipulation, furnishing a further check of our formalism.
Notice that setting $n=1$ here gives mean field exponents consistently  with the fact that {\tt Ising} has upper critical dimension $d_{c}=4$.
%As we will show later, the case of {\tt Ising} is the only one for which the upper critical dimension $d_c\neq 6$ and therefore this analysis is qualitatively wrong.
%For $n\geq2$ we must have $3n>7$ in order to have $\nu>\frac{1}{2}$, thus $n\geq3$ and $n=2$ is excluded implying some how that it is first order below six dimensions.
%
The explicit results for {\tt Percolation} are \cite{DeAlcantara1981, Bunde}
\begin{equation}
\nu=\frac{1}{2}+\frac{5}{84}\epsilon+O(\epsilon^{2})\qquad\qquad\eta=-\frac{\epsilon}{21}+O(\epsilon^{2})\,,
\end{equation}
while for {\tt Spanning Forest} we find \cite{Sokal2007}
\begin{equation}
\nu=\frac{1}{2}+\frac{1}{12}\epsilon+O(\epsilon^{2})\qquad\qquad\eta=-\frac{\epsilon}{15}+O(\epsilon^{2})\,.
\end{equation}
Note that for $n\geq2$ we find $\nu<\nu_{{\rm MF}}$, even if one expects these values to be meaningless since in  $d\geq 3$ the phase transition is first order.
%
%$$
%\nu=\frac{1}{2}-\frac{5}{12}\epsilon+O(\epsilon^{2})\qquad\qquad\eta=\frac{\epsilon}{3}+O(\epsilon^{2})\,.
%$$
%%
%{\color{red} For {\tt Ising} we get from the betas
%%
%\begin{equation}
%\nu=\frac{1}{2}+\frac{\epsilon}{12}+O(\epsilon^{2})\qquad\qquad\eta=0+O(\epsilon^{2})\,.
%\end{equation}
%}
%In the following section we proceed considering the full general system of  beta functions of section \ref{sec:betan}.
%
Finally, we remark that the {\tt Ising} $\epsilon$-expansion can also be correctly recovered if one expands around $d=4-\epsilon$ the beta functions \eqref{betan1}.

\begin{figure}
\includegraphics[width=0.6\columnwidth]{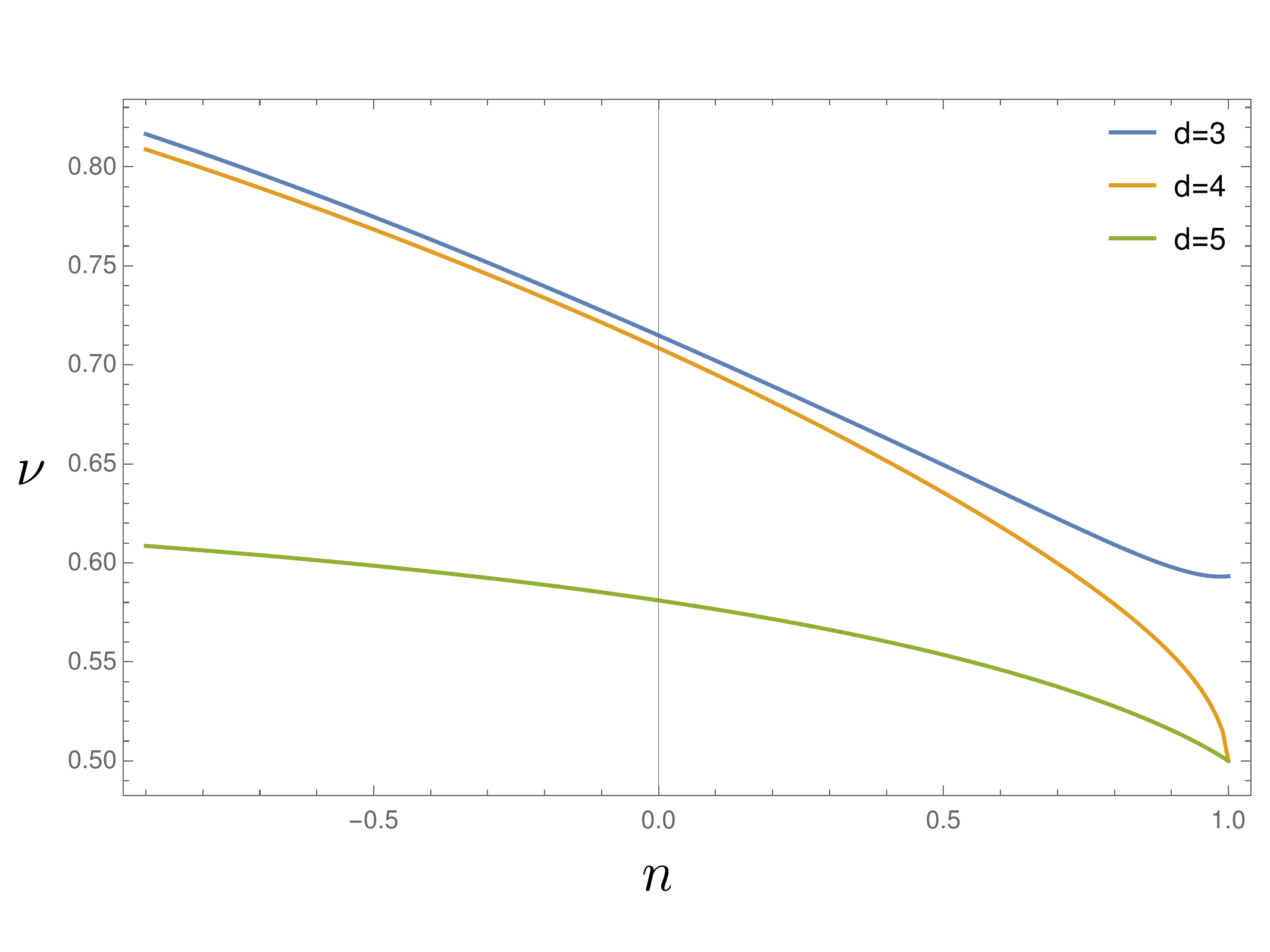}
\caption{The correlation length critical exponent $\nu$ as a function of $n$ for $d=3,4,5$ in the $\varphi^6$ truncation.
Curves interpolate smoothly between {\tt{Spanning Forest}} and {\tt{Ising}}: in $d=4,5$ the critical exponent converges to $\nu=\frac{1}{2}$ for $n=1$ since {\tt{Ising}} is mean field; in $d=3$ instead the curve approaches
$\nu=0.593$ which is the {\tt{Ising}} value at order $\varphi^6$ in the LPA'.
\label{Fig:nuUC}}
\end{figure}

\subsection{General Analysis\label{GA}}

The full set of beta functions presented in section \ref{sec:betan} can be studied only numerically. 
%Here we present numerical results for $p=6$ which corresponds to $N(6)=10$ couplings of non-derivative operators. 
As already explained, we focus on the non-trivial universality classes in $d>2$ which, apart {\tt Ising}, are {\tt Percolation} and {\tt Spanning Forest}.
The numerical analysis proceeds in steps: for all values $3 \leq p \leq 6$ we first solve numerically the algebraic system $\beta_{i,m}=0$ with $i=2,...,p$ and $m=1,...,N(p)$ to extract the fixed point coordinates;  we compute the stability matrix  $M_{ij}=\frac{\partial \beta_i}{\partial \lambda_{j}}$ symbolically and then we evaluate it at the numerical fixed point; finally we extract the eigenvalues.
At each truncation order we find just one negative eigenvalue $\theta_{-}$ from which we can extract the correlation length critical exponent $\nu =-\theta_{-}^{-1}$; the first positive one $\theta_{+}$ gives instead the correction to scaling exponent $\omega  =\theta_{+}$; the anomalous dimension is computed from  \eqref{eq:andim}. At every order of the truncation $p$, we repeated the procedure for any $3\leq d \leq 6$; the results for the critical exponents in function of the dimension $\nu(d), \omega(d), \eta(d)$  for orders considered $p=3,4,5,6$ are shown in Figure~\ref{Fig:nuP} for the relevant  {\tt Percolation} and {\tt Spanning Forest} cases.

It is immediately clear from the plots that as the order of the truncation $p$ increases the curves for the critical exponents converge non-uniformly: for values  $d$ closer to the upper critical dimension $d_c = 6$ few orders suffice to obtain a stable estimate. For $\nu$ in $d=5$ already $p=4$ returns good estimates; in $d=4$ instead the maximum order available $p=6$ is barely enough; while in $d=3$ convergence is still far from being reached and an improvement of the truncation is needed\footnote{As a first guess one can interpolate the curves using the converging parts; the result is consistent with the estimates for $\nu$ even in $d=3$. This gives us a  reason to expect that larger truncations can deal also with this case.}. In Table \ref{tablenu} we report the precise numerical values for the correlation length critical exponent.
For $\omega$ the convergence is, as expected, slower than for $\nu$ and the numerical estimates are reported in Table \ref{numomega}, while the results for the anomalous dimension in Table \ref{numeta}.
The estimates for $\nu$ are quite satisfactory and testify the success of polynomial truncations in the determination of this exponent. $\omega$ requires an improvement of the truncation while $\eta$ is poor, this being a general trend in LPA' like expansions.
In particular, it is difficult to see how the anomalous dimension can change sign in lower dimensions, as expected from $d=3$ estimates and the exact results in $d=2$, thus questioning the validity of \eqref{eq:andim} below four dimensions. 

\begin{figure}
\includegraphics[width=0.6\columnwidth]{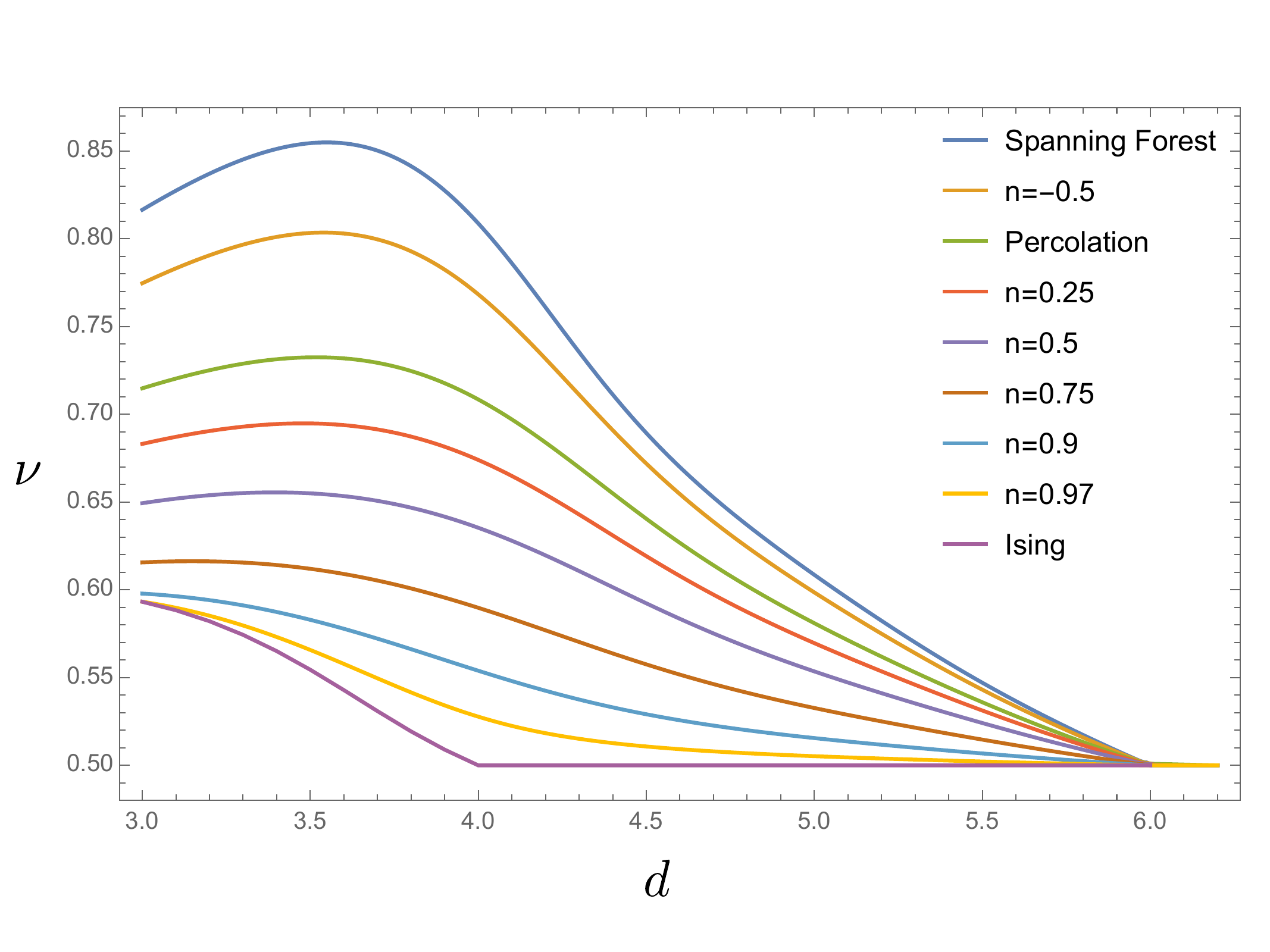}
\caption{Critical exponent $\nu$ as a function of $d$ for various values of $n$ in the $\varphi^6$ truncation. As $n$ increases, the critical exponent in the  $d>4$ region  flattens toward the $n=1$ mean field  value $\nu=\frac{1}{2}$ ({\tt Ising}).}
\label{Fig:RC}
\end{figure}

In Figure~\ref{Fig:nuUC} and  Figure~\ref{Fig:RC} we explore the $n$-dependence of the correlation length critical exponent. In the first figure we plot $\nu(n)$ for $d=3,4,5$ in the range $-1\leq n \leq 1$; in the second picture we plot $\nu(d)$ for values of $n$ between $-1\leq n \leq 1$ to see how the shift  of upper critical dimension from $d_c=6$ to $d_c=4$ when $n$ approaches one.

As shown in section \ref{PFT}, the number of couplings grows rapidly with the order $p$ of the truncation and even if the general analysis just presented indicate that larger truncations have the strength to fully determine the spectrum down to three dimensions, an explicit derivation of beta functions beyond those reported in section  \ref{sec:betan} demands a significant amount of further work. 
Finally, while in $d=5$ a study of scheme dependence is possible already at $p=6$ since convergence has been achieved, we postpone this study to a future work when also convergence in the other two physically relevant dimensions $d=3,4$ is obtained.

\begin{table}[t]
\begin{centering}
\begin{tabular}{cccccccc}
\hline 
dim & $\quad n\quad$ & $\quad\varphi^{3}\quad$ & $\quad\varphi^{4}\quad$ & $\quad\varphi^{5}\quad$ & $\quad\varphi^{6}$\quad &
\quad  best & \quad ref\tabularnewline
\hline 
\hline 
$5$ & $-1$ & $0.5481$ & $0.6059$ & $0.6071$ & \textcolor{blue}{0.60}85 &\quad  0.59&\quad \cite{Sokal2007}\tabularnewline
\hline 
 & $0$ & $0.5358$ & $0.5777$ & $0.5820$ & \textcolor{blue}{0.58}10 &\quad  $0.575$&\quad \cite{Gracey}\tabularnewline
\hline 
$4$ & $-1$ & $0.5551$ & $0.6492$ & $0.7476$ &  \textcolor{blue}{0.8}087 & \quad 0.80&\quad \cite{Sokal2007}\tabularnewline
\hline 
 & $0$ & 0.5415 & $0.6083$ & $0.6698$ &  \textcolor{blue}{0.7}084 & \quad $0.692$& \quad\cite{Gracey}\tabularnewline
\hline 
$3$ & $-1$ & 0.5468 & 0.6238 & $0.7151$ & $0.8170$ &  \quad$1.28$&\quad \cite{Sokal2007}\tabularnewline
\hline 
 & 0 & 0.5357 & 0.5927 & 0.6537 & 0.7148 & \quad 0.897& \quad\cite{Gracey}\tabularnewline
\hline 
\end{tabular}
\par\end{centering}
\caption{Correlation length critical exponent $\nu$ for {\tt Percolation} ($n=0$) and {\tt Spanning Forest} ($n=-1$) in $d=5,4,3$.
Estimates obtained in the various truncations considered are presented and convergent digits are denoted in blue. Comparison is made with available Monte Carlo simulations or re-summed high order $\epsilon$-expansion estimates.\label{tablenu}}
\end{table}

\begin{table}[t]
\begin{centering}
\begin{tabular}{ccccccccc}
\hline 
dim & $\quad n\quad$ & $\varphi^{3}$ & $\varphi^{4}$ & $\varphi^{5}$ & $\varphi^{6}$ %&  LPA
& best & ref\tabularnewline
\hline 
%$3$ & -1 & 3.827 & 4.458 & 4.488 & 4.270 & $?$\tabularnewline
%\hline 
% & 0 & 3.616 & 3.975 & complex & complex & 2.23\tabularnewline
%\hline 
\hline
5 & -1 & 1.157 & 0.8413 & 0.603 & \textcolor{blue}{0.6}39 %&\textcolor{blue}{0.639}
&  &   \tabularnewline
\hline 
 & $0$ & 1.116 & 0.842 & 0.627 & \textcolor{blue}{0.6}45 %& \textcolor{blue}{0.645}
 & 0.718 & \cite{Gracey}\tabularnewline
 \hline 
 $4$ & -$1$ & 2.466 & 2.500 & 1.969 & 1.368 %&\textcolor{blue}{1.368} 
 &  & \tabularnewline
\hline 
 & $0$ & 2.346 & 2.295 & 1.922 & 1.587 %&\textcolor{blue}{1.587}
 & $1.2198$ & \cite{Gracey} \tabularnewline 
\hline 
\end{tabular}
\par\end{centering}
\caption{
Critical exponent $\omega$ for {\tt Percolation} ($n=0$) and {\tt Spanning Forest} ($n=-1$) in $d=5,4$.
Estimates obtained in the various truncations considered are presented and convergent digits (in the LPA') are denoted in blue. Comparison is made with available re-summed high order $\epsilon$-expansion estimates in the $n=0$ case. 
No estimates have been found in the literature for $n=-1$.\label{numomega}
}
\end{table}

\begin{table}[t]
\begin{centering}
\begin{tabular}{cccccccc}
\hline 
dim & $\quad n\quad$ & $\varphi^{3}$ & $\varphi^{4}$ & $\varphi^{5}$ & $\varphi^{6}$ & best & ref\tabularnewline
\hline 
\hline 
%$3$ & -1 & 1.184 & 1.647 & 2.257 & \textcolor{blue}{3.003} & $2.77$\tabularnewline
%\hline 
% & 0 & 1.138 & 1.426 & 1.747 & \textcolor{blue}{2.058} & 1.805\tabularnewline
%\hline 
$5$ & -$1$ & -0.055 & -0.145? & -0.145? & \textcolor{blue}{-0.145}8 &  -0.08 & \cite{Sokal2007} \tabularnewline
\hline 
 & $0$ & -0.041 & -0.099 & -0.104 & \textcolor{blue}{-0.10}2 &   -0.0565 & \cite{Gracey} \tabularnewline
\hline 
4 & -1 & -0.110 & -0.382 & -0.590 & -0.678 &  -0.16&\cite{Sokal2007}\tabularnewline
\hline 
 & $0$ & -0.0833 & -0.250 & -0.363 & -0.406 &   -0.0954&\cite{Gracey}\tabularnewline
\hline 
\end{tabular}
\par\end{centering}
\caption{
Anomalous dimension $\eta$ for {\tt Percolation} ($n=0$) and {\tt Spanning Forest} ($n=-1$) in $d=5,4$.
Estimates obtained in the various truncations considered are presented and convergent digits (in the LPA') are denoted in blue. Comparison is made with available Monte Carlo simulations or re-summed high order $\epsilon$-expansion.
As generally happens with LPA' truncations, the estimates for the anomalous dimension are much poorer that those for $\nu$ or $\omega$. \label{numeta}
}
\end{table}

%\subsubsection{Comment on convergence}
%
%In the LPA scheme of section (?) we already checked the functional
%form of the $\beta$-functions in the limit $n\to1$ truncating the
%potential to $N_{\text{tr}}=3$. One way to improve the approximation
%stands for truncating the potential to a polynomial of higher order.
%We studied the critical exponents in $3\leq d\leq4$ truncating the
%potential at $N_{\text{tr}}=6$ where we consider $\varphi^{12}$
%interactions. Figure (?) shows the critical exponents $\nu$ in the
%critical range: we see that truncating the potential at higher orders
%amounts to sistematically improve the numerical value of the critical
%exponent which is converging to the value $\nu_{\text{LPA}}(3)=0.651$
%to be compared with $\nu_{\text{best}}(3)=0.629$.
%
%\begin{figure}[h]
%\includegraphics[width=0.7\textwidth]{Graphics/Ising.pdf}
%\caption{We keep this to comment about LPA convergence of critical exponents KEYNOTE!}
%\end{figure}
%
%The analysis was carried on the LPA scheme with $\eta$=0 at all orders.
%Better estimates follow from higher oreder truncations in the LPA'
%scheme where the anomalous dimension is included {[}?{]}.

\section{Conclusion and Outlook}

%\subsubsection{What we achived}

The main goal of the present paper, after an initial discussion of the Potts field theory, with particular attention to the construction and enumeration of (non-derivative) invariants, has been the adaptation of functional RG (FRG) methods to the field theory of an $n$-component scalar with the underlying symmetry of the Potts model: global $S_{n+1}$.

Our main result has been the development of an algorithm able to compute the beta functions for the couplings of potential
interactions, which we then used to perform explicit computations up to order $\varphi^{6}$, thus obtaining a system of coupled ODE describing the RG flow of the ten couplings present at this order, for arbitrary $d$ and $n$.
The main technical difficulty in the application of FRG methods to $S_{n+1}$-symmetry consisted in a systematic enumeration and construction of the invariants and in the development of the necessary trace machinery to reduce the traces present in the expansion of the r.h.s. of the flow equation, unlocking in this way the access to the beta functions. 

While it is not possible to obtain a closed equation for the effective potential for arbitrary $n$,
it is instead possible to do so for specific integer values. We derived the improved local potential approximation (LPA') explicitly for the $n=1,2$ cases and explained how to do
it for arbitrary positive integer $n$. A characterising property of the $S_{n+1}$-symmetry is that the LPA' is a PDE of $(n+1)$-variables, since $n$ independent invariants can be built out of the field multiplet even without introducing derivatives.
In this respect the generalization of a single component $\mathbb{Z}_2$-scalar to a multi-component scalar is more involved in the $S_{n+1}$ case than, for example, in the $O(n)$ case.

As a first application, after a simple study of a two coupling truncation comprising mass and $\varphi^3$-coupling, which also allowed us to recover the leading order $\epsilon$-expansion, we obtained
% (using a ten coupling truncation with operators up to order $\varphi^6$)
estimates for the critical exponents for the {\tt Percolation} and {\tt Spanning Forest} universality classes in dimension $d=4,5$ (where a preliminary study of convergence was possible) and made a first analysis in $d=3$.
Our numerical estimates for $\nu$ and $\omega$ turned out to be in quite satisfactory agreement with Monte Carlo simulations and high order $\epsilon$-expansion results, showing that FRG methods are indeed very effective also in the case of $S_{n+1}$-symmetry.

%\subsubsection{Problems to be solved in the future}

Apart for these first results and applications, our study serves as groundwork for
future enquires of the {\tt Potts}$_{n+1}$ universality classes.
Several questions demand further study or remain unanswered, the main of which are:
\begin{enumerate}

\item Study larger truncations to achieve full convergence in $d=3,4$ and study regulator dependence to extract the best estimates for the critical exponents.

\item We recall the problem of the sign change in $\eta$ from a negative
value in $d=6-\epsilon$ to a positive one in $d=2$.
To overcome the problem, and to obtain better estimates for $\eta$, one must probably consider truncations that go beyond  non-derivative interactions (i.e. beyond LPA').

\item One needs to study the two variable PDE encoding the flow in the LPA' for the three states Potts
model ($n=2$) to understand for which value of $d$ the phase transitions becomes first order.
This value is expected to lie between two and three \cite{Nienhuis1979}. A similar study could be made for the three variable PDE pertaining to the $n=3$ case.

\item  Much less is known about possible multi-critical
phases of  models with $S_{n+1}$-symmetry in any dimension grater than two, and for which more understanding is needed. An analysis of these phases can be done with LPA' scaling solutions following \cite{Codelloscaling1, Codelloscaling2, Codelloscaling3, zambellizanusso}.

\item Obtain the LPA' for continuous values of $n$ in order to enable a fully functional analysis of the physically
interesting limits $n\to0$ and $n\to-1$.
A large-$n$ limit will also become available. 

\end{enumerate}
As stated in the introduction, this paper is the first of a series devoted to the study of universality classes characterized by discrete global symmetries in arbitrary dimension \cite{bencodello2}.
We will also be interested in applying the complementary methods of functional perturbative RG \cite{Codello1} and CFT+SDE \cite{Codello0} to Potts field theories in order to obtain their spectrum and OPE coefficients in the $\epsilon$-expansion.
Multi-critical models with $S_{n+1}$-symmetry can also be explored along the lines of the relative one component {\tt Lee-Yang} family \cite{Codello2} and may represent non-trivial interacting theories $d=3$.
Another line of future developments is to study long range interactions \cite{CodelloDefenuTrombettoni} with Potts symmetry, as recently done for {\tt Percolation} in \cite{TrombettoniDefenuGori}.
Otherwise one can study systems out of equilibrium \cite{PottsOOE, Canet} generalizing the work of \cite{Chiocchetta} or consider the presence of boundaries \cite{Diehl}.

\section{Acknowledgments}
R.B.A.Z.  would like to thank Giacomo Gori and Patrick Azaria for interesting discussions and useful comments.

\end{document}